Date of acceptance Grade

Instructors
Jürgen Münch
Tomi Männistö

# A Systematic Approach to Setting Up Distributed Global Collaborations for Software-based Products in the Automotive Domain

Inna Smirnova

Helsinki August 15, 2014

UNIVERSITY OF HELSINKI
Department of Computer Science

HELSINGIN YLIOPISTO – HELSINGFORS UNIVERSITET – UNIVERSITY OF HELSINKI

| Tiedekunta/Osasto – Fakultet/Sektion – Faculty/Section | Laitos – Institution – Department |
|---|---|
| Faculty of Science | Department of Computer Science |

Tekijä – Författare – Author

Inna Smirnova

Työn nimi – Arbetets titel – Title

A Systematic Approach to Setting Up Distributed Global Collaborations for Software-based Products in the Automotive Domain

Oppiaine – Läroämne – Subject

Computer science

| Työn laji – Arbetets art – Level | Aika – Datum – Month and year | Sivumäärä – Sidoantal – Number of pages |
|---|---|---|
| Master’s thesis | August 15, 2014 | 72 pages |

Tiivistelmä – Referat – Abstract

There is an increasing need for organizations to collaborate with internal and external partners on a global scale for creating software-based products and services. Many aspects and risks need to be addressed when setting up such global collaborations. Different types of collaborations such as engineering collaborations or innovation-focused collaborations need to be considered. Further aspects such as cultural and social aspects, coordination, infrastructure, organizational change process, and communication issues need to be examined. Although there are already experiences available with respect to setting up global collaborations, they are mainly focusing on certain specific areas. An overall holistic approach that guides companies in systematically setting up global collaborations for software-based products is widely missing.

The goal of this thesis is to analyze existing literature and related information and to extract topics that need be taken into account while establishing global software development collaborations - to identify solutions, risks, success factors, strategies, good experiences as well as good examples. This information is structured in a way so that it can be used by companies as a well-grounded holistic approach to guide companies effectively in setting up long-term global collaborations in the domain “software development”. The presented approach is based on scientific findings reported in literature, driven by industry needs, and confirmed by industry experts.

The content of the thesis consists of two main parts: In the first part a literature study analyzes existing experience reports, case studies and other available literature in order to identify what aspects and practices need to be considered by organizations when setting up global collaborations in the domain software development. Secondly, based on the results from the literature review and consultation with the industrial partner Daimler AG, the identified aspects and practices are structured and prioritized in the form of activity roadmaps, which present a holistic guide for setting up global collaborations. The developed guidance worksheet, the so-called “Global canvas”, is meant to be a guide and reminder of all major activities that are necessary to perform when doing global collaborations for software-based products and services.

The main contributions of this thesis are an analysis of the state of the practice in setting-up of global software development collaborations, identification of aspects and successful practices that need to be addressed by organizations when doing global collaborations for software-based products and services and the creation of a holistic approach that presents scientific findings to industry in an effective and credible way and guides companies in systematically setting up global collaborations.

# Contents

# 1 Introduction

## 1.1 Global collaborations and organizations' needs

*"Recognized as one of the trends of the 21st century, globalization of the world economies brought significant changes to nearly all industries, and in particular it includes software development. Many companies started global software engineering (GSE) to benefit from cheaper, faster and better development of software systems, products and services"* [SWG10]. Nowadays the Global Software Development (GSD) with process distribution all over the world is seen as "a normal way of doing things" [SWG10]. Through global collaborations companies expect potential benefits such as access to low-cost resources, particularly a labor pool with diverse expertise and skills that could easily scale-up software development teams and potentially lead to financial savings [FBD08]. Other benefits for companies might be the possibility to speed up the development process, freeing up local resources, accessing foreign know-how, tapping into new markets and customers and therefore achieving the desired market competitiveness [AFH05, KRF01, LLA07, SWA13, Smi06, FBD08, Rot06].

Some companies already work with Global Software Development collaborations, while other clearly see the need and potential benefits and thereby intend to start setting up global collaborations [SWG10]. However, along with potential benefits Global Software Development might bring many risks and challenges such as communication, culture differences, coordination, and project management efforts [HPB05]. This means that practitioners who have yet to achieve a certain level of experience in setting up global collaborations need to consider many aspects, challenges, different strategies, available solutions and experiences. Such knowledge might help to reduce possible future negative effects leading to GSD project failures such as cost overruns, timeframes exceeding, low product quality and overall decreased customer satisfaction [SWG10, NJS11, PAD07]. Organizations need to adapt their current practices of software development in order to benefit and gain competitive advantage in the global context [RCM12].

The challenge for organizations that intend to set-up global collaborations in the domain software development therefore is depending on the following: there is a need for a holistic approach when setting up global collaborations which aggregates available practices, experiences, solutions and alternatives and guides companies in systematically setting global collaborations for software-based products and services. Organizations need to be aware of the aspects that necessarily have to be considered when setting up global collaborations. Such a holistic approach is widely missing and this is a problem [BOR13]. This thesis aims at closing this research gap and creating a holistic approach for setting up global collaborations in the software development domain.

## 1.2 Thesis goals

**The goal of this thesis is the development of an overall holistic approach that guides companies effectively in setting up long-term global collaborations in the domain software development. The approach will be based on scientific findings reported in literature, driven by industry needs, and validated through industry experts.**

Systematic literature study will be conducted as a main research method used for extracting data needed for the following guide creation. Moreover, the research will be supported by the cooperation with the industrial partner company that operates in the automotive domain. Based on the literature review and advices from the industry, the principal aspects that need to be considered when setting up global collaborations will be investigated. Appropriate practices for doing GSD collaborations based on available literature and experiences will be extracted. Afterwards, discovered aspects and practices will be prioritized and structured in the form of activity roadmaps that aim at a synthesis of all relevant experiences and could be helpful for practitioners intending to set-up global collaborations.

Based on the results the initial version of a worksheet, the so-called “Global canvas”, will be developed. The worksheet aims to function as a guidance and a reminder for companies on what major aspects and activities need to be addressed by organizations when setting up Global Software Development collaborations. The goal is to provide a worksheet that presents scientific findings to the industry in an effective way and helps to identify what needs to be considered when setting up global collaborations. The aim of this study is to come up with an initial proposal for such a prescriptive worksheet that is driven and validated against the needs and requirements of the case company. The worksheet creation is based mainly on the needs of the case company, and that a detailed validation of aspects and practices is out of scope in this thesis.

This thesis is intended to be continued with further research that is aimed to test the proposed guidance worksheet in the industrial conditions and investigate its practical value. In the future research the new versions of the “Global canvas” worksheet might be created and new potential ways of using the worksheet by organizations could be investigated.

The results of the thesis have been described in an article that is accepted for publication at the 20th International Conference on Information and Software Technologies (ICIST 2014) [SMS14]. Parts of this article were integrated into this thesis.

## 1.3 Overview of the thesis

Chapter 2 describes the related work. The existing research and experiences on establishing global collaborations in the software development domain will be discussed. The focus of the current research on global collaborations will be analyzed, particularly regarding its practical use for organizations.

In Chapter 3 the study context and the research method will be explained along with necessary steps performed for collecting data.

Chapter 4 presents the results from the conducted literature review and industry consultations. The chapter will describe in detail which principal aspects and practices for setting up global collaborations can be identified.

In Chapter 5 the worksheet “Global canvas” is developed as an aggregation of the literature study findings presented in Chapter 4. The prioritization of aspects and practices needed to be addressed by organizations when establishing GSD collaborations will be proposed. The visualization of the worksheet will be introduced and explained.

In Chapter 6 the discussion of the results and study limitations will be analyzed.

Chapter 7 concludes findings of this thesis and suggests potential for future work.

# 2 Related Work

A wide variety of studies already exists in the area of global software engineering that analyzes globally distributed projects, their risks and challenges with a comparison to traditional co-located software development project, and proposes some risk mitigation advices and experiences.

Nurdiani et al. performed a systematic literature review among GSD research literature that resulted in a checklist of 48 GSD challenges and 42 mitigation recommendations [NJS11]. The challenges and mitigation strategies were based on the impact of temporal, geographical and socio-cultural distances which appear in globally distributed collaborations on the communication, coordination and control aspects.

Another systematic literature review was done by Verner et al., who reported the risks of GSD collaborations with some mitigation recommendations structured into 12 areas, starting from vendor selection, requirements engineering and finishing with coordination and control areas [VBK14]. This research presents explicit tables of potential GSD risks, but mitigation advices are quite limited and not described in detail or not presented at all for some risks.

Šmite et al. conducted a systematic literature review of GSD experiences and came up with the seven most commonly discussed practices that are aimed at overcoming GSD problems [SWG10]. The identified risk mitigation practices are presented with their benefits and difficulties to conduct which might be helpful for practitioners to look at first, before actually implementing suggestions.

Mettovaara et al. performed interviews at Nokia and Philips and identified 10 common problems and 11 success factors based on experiences in interorganizational collaborations in the two studied companies [MSL06].

However, all those above-mentioned studies are first and foremost risk- and problem-oriented. The thesis study takes these findings into account. But instead of identifying relevant risks, it aims at providing a constructive guide that contains helpful practices and a sequence of activities for setting up global collaborations in the software development domain [SMS14].

Also, many studies on GSD topic were identified to be dedicated to some specific aspect of global collaborations such as communication or trust.

Paasivaara and Lassenius conducted semistructured interviews at six Finnish companies in order to examine collaboration practices with a special focus on communication that proved to be successful in real global interorganizational software development projects [PaL03]. The relationship building and team awareness channels as well as the synchronization of main process milestones between collaboration sites were considered to be crucial parts of communication in global collaborations that impact how successfully the software development project will be completed.

Thissen et al. [TPB07] and and Niinimäki et al. [NPL10] performed case studies in order to analyze what variety of communication tools are mostly used for distant remote communication and information exchange between globally distributed teams in GSD projects. Studies analyzed the use of such remote communication ways as instant messaging tools, E-mails, phone and video conference tools along with their potential advantages and suitability for specific information types.

Piri et al. [PNL12], Pyysiäinen [Pyy03] and Moe et al. [MoS08] performed interview-based studies to investigate in detail the role of trust in global projects, its impact on global collaborations and supportive successful practices for trust building. The reasons for lack trust in global collaborations, main negative effects of lacking trust, trust building phases and recommendations were examined. Trust is claimed to be one of the key success factors that guarantee building longitude, successful and reliable global collaborations. Therefore trust building practices are recommended to be performed and maintained by organizations through the whole collaboration history.

Oshri et al. [OKW08] investigated the aspect of building social ties between sites in global collaborations. The special emphasis of the study was given to face-to-face meetings as a main solution for building collaborative social ties between distributed software development teams. The necessary activities before, during and after face-to-face meetings were presented. The social ties between collaboration sites were viewed at individual, team and organizational levels.

Abraham [Abr09], Huang et al. [HuT08], MacGregor et al. [MHK05] and Winkler et al. [WDH08] examined the impact of cultural distance on global collaborations. Studies investigated how cultural distance affects the work performance by individuals, their communication and behavioral style. Cultural diversity has an impact on trust, communication and relationship building between collaboration sites. Studies conclude that cultural understanding is crucial for building cooperative global collaborations.

Šmite conducted a case study at a Latvian software company and reported challenges and the 8 practices used regarding the coordination and project management aspects [Smi05]. Study also proposes possible work distribution and division of development process activities between collaboration sites.

Hossain et al. [HBV09] conducted a case study on Australian-Malaysian cooperation that examines the alternative ways of coordination mechanisms in globally distributed projects. Three coordination mechanisms – mutual adjustment, direct supervision and standardization were described. The impact of temporal, geographic and socio-cultural distances on coordination mechanisms was studied. Some success practices of performing coordination were introduced based on their testing in the studied collaboration.

Mockus et al. [MoW01], Salger [Sal09] and Lamersdorf et al. [LMR08] examined task allocation between sites in Global Software Development collaborations. The approaches for transferring work to remote locations were presented as, for example, transfer work by functionality, by localization need, or by development stage. The necessary elements of work packages for transfer were introduced. A work package, for example, can consist of software requirements specifications, design artifacts and project management artifacts.

There are also some studies focusing on GSD topic that try to create an overall checklist of aspects needed to be considered by organizations when doing global collaborations.

Šmite et al. [SWA13] performed a literature study along with a case study at Ericsson and presented decision-making process guidance in form of a checklist of questions which need to be considered when setting up global collaborations. Their focus can be considered to be mostly on initiating and planning phases of collaborations so that the resulting guide considers such aspects as - Why do we do global collaboration? For what type of work? When, where and how do we divide work, roles and responsibilities? [SWA13]. The developed guide can be seen as a high level decision making checklist and it does not consider all specific collaboration aspects and successful practices in detail. In contrast, this thesis study aims at covering all aspects that need to be considered when establishing global collaborations in the domain software development, including not only initiating and planning phases, but also operating and improvement. Moreover, it aims at finding an order of activities that need to be done when setting up global collaborations.

Richardson et al. conducted literature studies and empirical case studies within the industry and created a Global Teaming process framework that contains threats, practices and guidelines for implementing global cooperations in the industry [RCM12]. Although there are some similarities with the thesis goals, the thesis aims at discovering whether there are different aspects and practices that need to be addressed by organizations. Moreover, the thesis aims at prioritizing activities and creating a roadmap for companies by dividing activities into collaboration stages.

Overall, the research topic of Global Software Development collaboration has a wide interest in literature. A lot of scientific studies that examine different aspects of GSD such as its potential benefits for organizations, new risks and challenges, possible strategies of organizing global collaborations, and successful practices that mitigate the impact of global distance risks already exist. There are also industrial studies that report different case studies on GSD collaborations in order to discover what could work in reality, what needs to be improved, and what models of global collaborations are preferred by organizations.

However, in general, the related work shows that literature regarding the set-up of global collaborations in the domain software development is either risk-oriented or narrow with a focus on a specific aspect of collaboration such as communication, trust building, coordination mechanisms, or task allocation. There are some approaches that give a holistic view on practices and guidelines for executing global software development collaborations. However, some mainly put the risks of GSD collaborations in the first place and try to find mitigation solutions, while some focus on successful practices in a form of high level checklists that are hardly of practical use. Existing literature studies in the Global Software Engineering domain do not focus on activity prioritization when setting up global collaborations and therefore might not give enough guidance for industrial companies. The analysis of related work offered motivation for this thesis to create a synthesis of relevant aspects and practices structured in a form of a guide for companies that intend to set-up global collaborations and need an overall high level view of activities that need to be considered.

# 3 Context and Research Method

The study was performed in collaboration with an automotive OEM “Daimler AG” that served as a case company for this study [SMS14]. The respective business unit of the company that was the contact point for this study is intending to set-up a long-term, multi-national distributed global collaboration for software-based products and services in the automotive domain [SMS14]. Based on the experiences from previous projects, the company identified a set of aspects that were seen as highly important when setting up global collaborations. The aspects included, for instance, collaboration structure, product structure, communication, organizational change process. Furthermore, the company provided an order in which the aspects are proposed to be addressed when doing global collaborations. The information about the aspects and their order was gathered from the experience from project leaders in the global projects within the business units “Daimler trucks” and “Daimler buses” [SMS14]. In addition, members of case company attended several ICGSE (International Conference on Global Software Engineering) conferences and input from these conferences implicitly influenced the selection of the aspects [SMS14].

The aspects provided by the case company are used in this thesis as means for structuring the areas with practices for setting up global collaborations in the software development domain [SMS14]. This was the main rationale for selecting the aspects. Performing a systematic literature study was chosen as a main technique for collecting necessary data in order to identify the principal practices for aspects that needed to be addressed by organizations when setting up global collaborations.

Snowballing was chosen as a main research method for the required literature review [WeW02, BFM13, Woh14]. It was also chosen as an instance of systematic approach to literature review that helped collect all the necessary relevant literature without performing a full Systematic Literature Review (SLR) [SMS14]. The topic of research interest was very broad at the beginning, therefore it was decided not to perform a full SLR, but to choose a different systematic approach. Snowballing was found to be suitable for this exploratory research [SMS14].

Based on Webster and Watson [WeW02] as well as Wohlin [Woh14] the starting point for the snowballing research approach is to analyze the main contributions to the research topic, in order to identify a set of key papers that could be used in further steps of the snowballing procedure. The main issues for the key papers are the topic relevance and the diversity [Woh14]. As a starting point in this study it was decided to analyze several key systematic literature reviews related to setting up GSD collaborations. Four key SLRs research studies were selected - [SWG12, NJS11, VBK14, SWG10]. They aggregate the experience on risks, mitigation solutions, and strategies regarding GSD collaborations. These key SLRs provide an explicit overview of related work in the area of Global Software Development and give a good input on the dimensions of global collaborations that might be important to address by organizations. They were therefore seen as a suitable starting point.

The next step in the snowballing method according to Wohlin [Woh14] is the combination of backward and forward snowballing. Backward snowballing requires reviewing the reference lists in the papers that serve as the starting point with the goal to identify topic-relevant studies that need to be considered [WeW02, BFM13, Woh14]. The candidate papers in the reference lists are examined with respect to authors, titles and publication venues. Moreover, the relevance of the candidate papers is evaluated by reviewing where and how they are referenced in the papers that serve as the starting point. Forward snowballing requires analyzing the citations to the key papers that are examined. The relevance and suitability of the candidate papers in the forward snowballing is evaluated in the same way as in the backward snowballing. In this step a review of the bibliographic reference lists of the four SLRs and the citations to them was performed. The studies that fit to the aspects defined by the case company were selected [SMS14]. This snowballing step then was iteratively performed up to four times depending on the suitability of the results found [SMS14]. Figure 1 summarizes the performed Snowballing procedure.

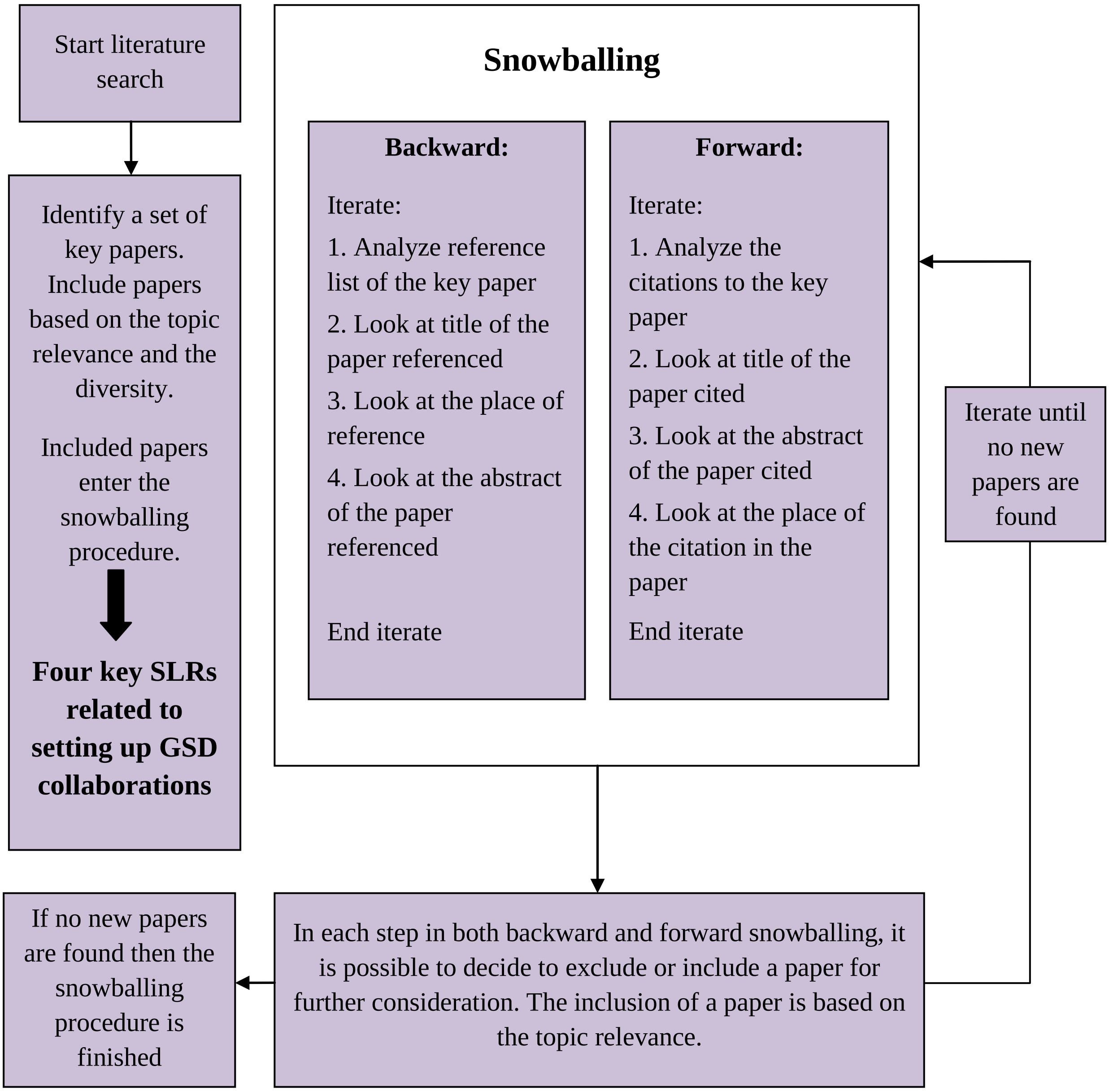

Figure 1. Snowballing procedure [Woh14]

In order to identify more topic-relevant papers, scientific papers from major conferences on global collaborations were systematically reviewed additionally. Two conferences were chosen as promising sources for collecting relevant data - International Conference on Global Software Engineering (ICGSE, reviewed proceedings for a period of 2006-2013) and International Conference on Software Engineering Approaches For Offshore and Outsourced Development (SEAFOOD, reviewed proceedings for a period of 2007-2010).

After the literature search, the content relevance of selected papers was examined in detail. The most promising were defined as a literature pool for the current master thesis. The next step of the study represented data extraction from the found literature. It was followed by further analysis regarding the detailed specification of aspects, identification of strategies, and practices that need to be addressed in global software development projects [SMS14]. Following Whittemore and Knafl [WhK05], the gathered aspects and practices were grouped together in an integrative way, divided into different phases of global collaborations and prioritized [SMS14]. As a result a guidance worksheet, the so-called “Global canvas” which aims at supporting practitioners for setting up global software development collaborations, was created. As an initial validation, the results were frequently presented to the key stakeholders in the case company and reviewed [SMS14]. Feedback was used to revise the worksheet “Global canvas” and create the final version presented in the thesis [SMS14]. Figure 2 summarizes the overall research process performed in this thesis.

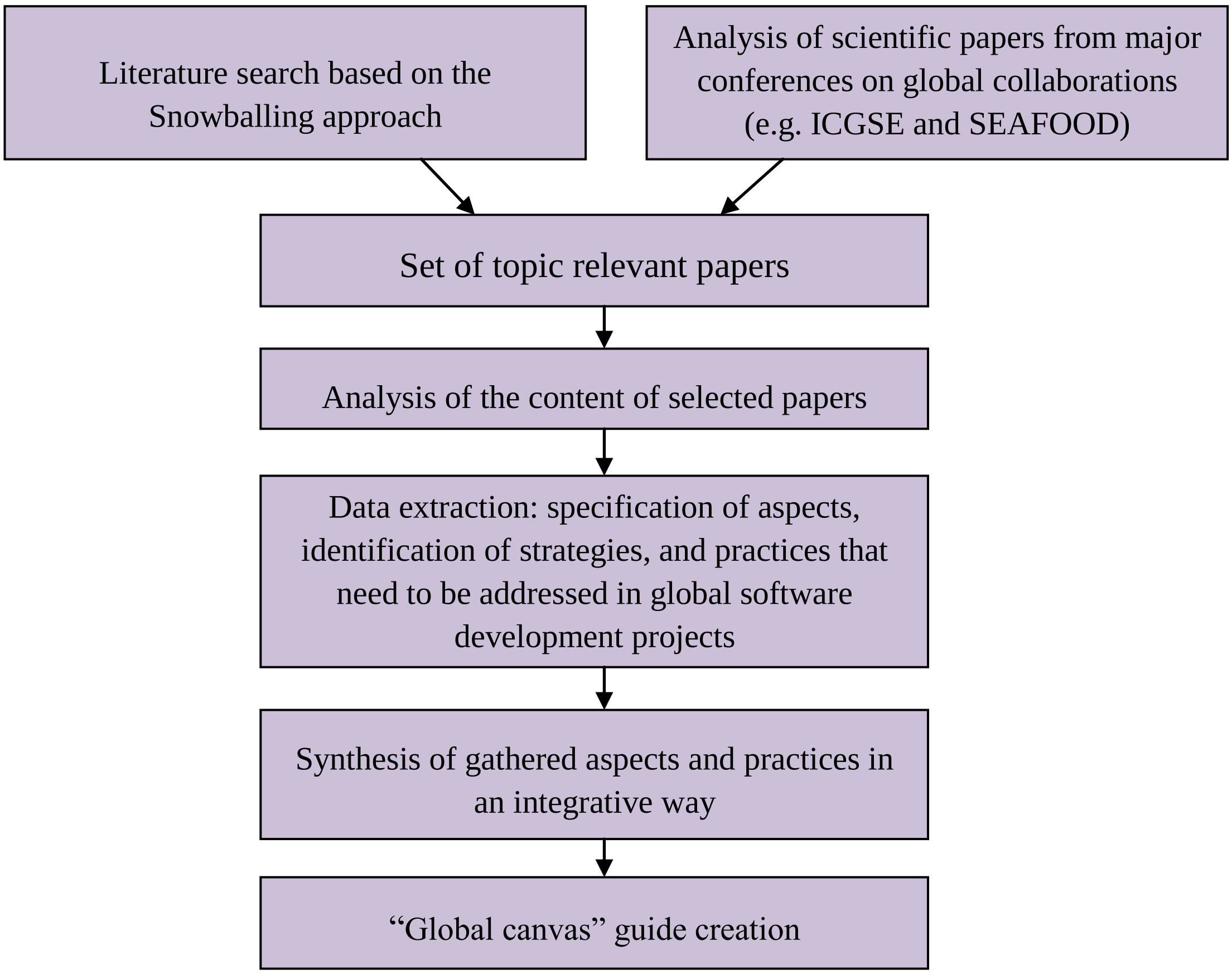

Figure 2. Research process steps: overview

# 4 Results from the Literature Study

This chapter will introduce the results from the performed literature study and industry consultation. The study was performed in collaboration with the case company that operates in the automotive domain and intends to set-up a long-term global collaboration for software-based products and services [SMS14]. The case company provided a set of aspects that were considered as highly important when working with global software development collaborations [SMS14]. These aspects are used in the study as a main rationale for structuring the practices and experiences reported in the literature [SMS14]. Therefore the practices and experiences are structured into nine identified main aspects that are proposed to be considered by organizations when setting up global collaborations in the software development domain. Based on the analysis of existing literature, particularly Systematic Literature Reviews related to global collaborations, the selected main aspects were slightly adjusted, especially regarding their definition and naming. This chapter presents the detailed description of the aspects along with their importance in global collaborations.

Each aspect consists of listed successful practices and extracted experiences that represent the set of important activities suggested to be performed by organizations in order to set-up successful global collaborations. Each practice will be described in detail according to the following structure (Figure 3):

- **Objectives** – What are the goals of the practice and why is it important in GSD collaborations?
- **Potential results and experiences** – What can be achieved with the practice? What are the “good” and “critical” points? Which experiences exist?
- **Actions** – How to conduct the practice? What alternatives of practice activities exist?

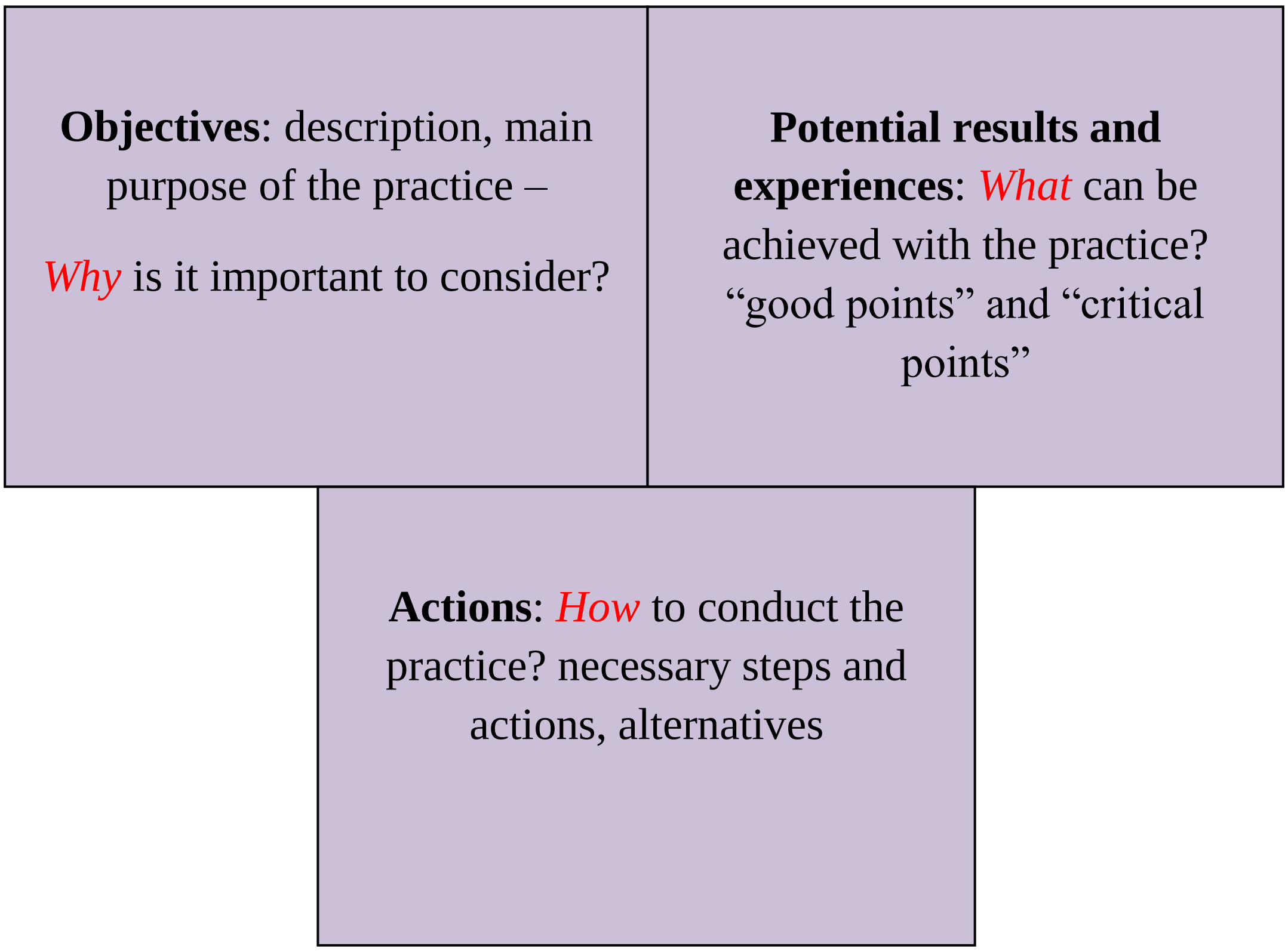

Figure 3. Structure of the practice

The rest of the chapter will present identified aspects and practices that need to be addressed when working with global collaborations.

## 4.1 Strategy

As it was previously mentioned in the Introduction chapter, nowadays many software companies understand the need to operate globally [SWG10]. Companies who intend to transfer software development work from a co-located context into a globally distributed model first need to analyze potential outcomes, reasons, existing strategies, and first steps of working with global collaborations [SWG12, SmW12].

The first identified aspect to be addressed by organizations when setting up global collaborations is *Strategy*. It aims at answering such questions as – Why do we collaborate globally? What are the potential benefits that transition into Global Software Development model might bring? How to perform global collaborations, according to what model? What collaboration partner to choose? [SMS14]. The answers for those questions represent the base for global collaborations, the first crucial steps, so that they cannot be avoided by organizations.

With the aspect of *Strategy* five main practices were extracted from the literature – define Collaboration goals and Collaboration model, investigate foreign legal system, choose vendors, and define budget plan [SMS14].

### 4.1.1 Collaboration goals

**Objectives:** The practice of defining Collaboration goals aims at identifying the reasons and potential outcomes that the setting-up of global collaborations might promise compared to traditional co-located software development.

**Potential results and experiences:** Different collaboration goals affect the whole high level framework of global collaborations – such as what model of collaboration could me more suitable for specific organizational context, what vendors have required expertise and skills needed to fulfill defined collaboration goals, what budget constraints are appropriate for the specific project that is globally distributed.

Based on the industry surveys of Forbath et al. [FBD08] and Silverthorne [Sil07] the main potential drivers of doing global collaborations for organizations can be classified into three areas (Figure 4).

The primary driver for doing global collaborations is financial savings, the so called cost leadership [FBD08]. The development costs might decrease due to access to a labor pool in countries with lower wage rates or access to foreign resources that are expensive and hard to obtain onsite, but easily available at offshore destinations [FBD08].

Another potential that companies see in global collaborations might be the access to capability that could be unavailable onsite such as foreign know-how, expertise and new technologies [FBD08]. These potential outcomes can be considered as innovations for organizations and might promise product optimization.

Also, the proximity to new offshore markets can be a driver for global collaborations that aim at product localization and customization. The contextual knowledge of the market might help to create a new value for products that attracts new customers and leads to bigger revenues for companies [FBD08, Sil07, AFH05, KRF01, LLA07, SWA13, Smi06, Rot06].

Altogether, all the collaborations goals that organizations might have can be summarized into the desire to get valuable market competitiveness.

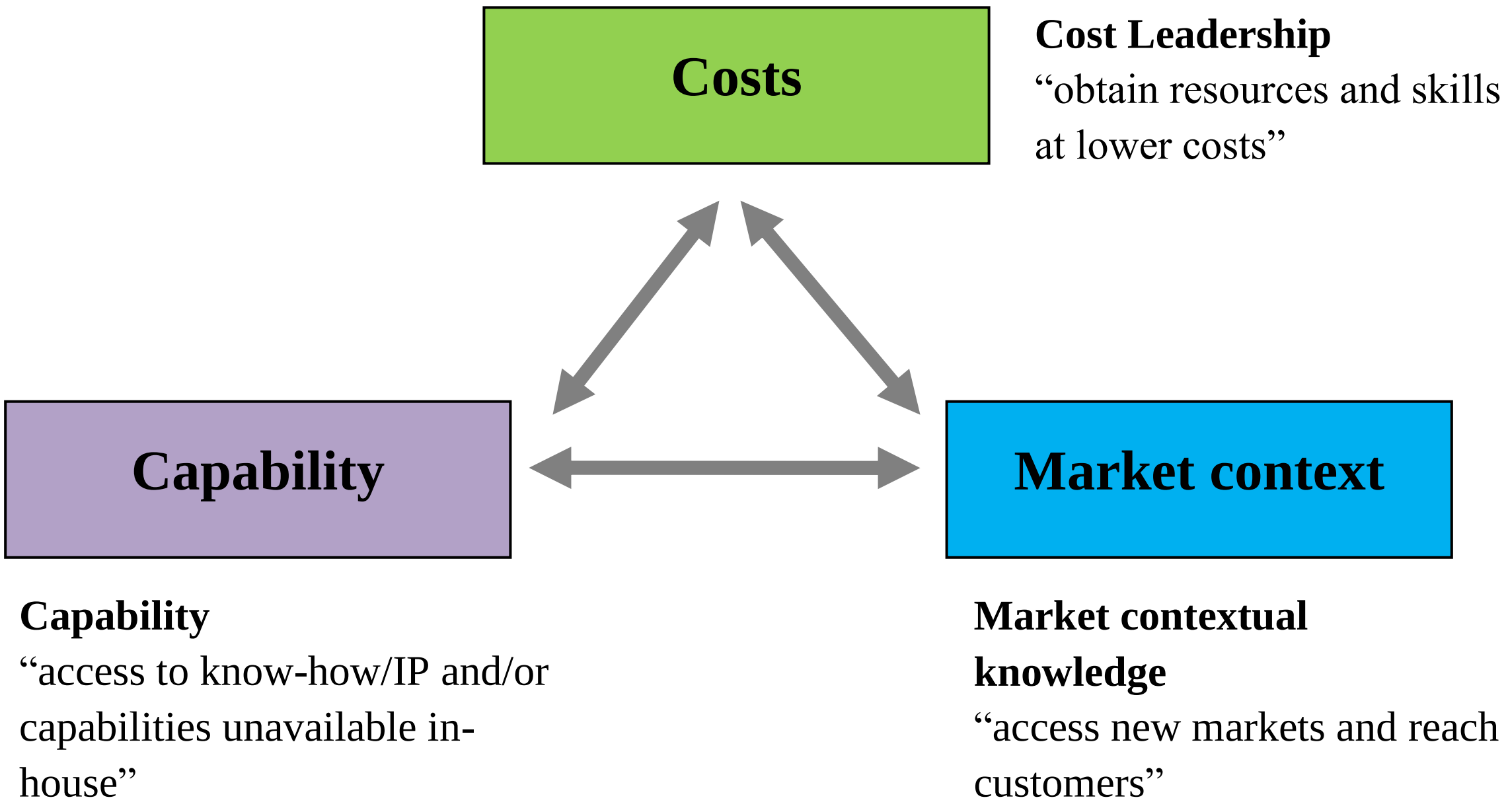

Figure 4. Drivers of global collaborations [FBD08]

**Actions:** Choose correct primary goals for performing global collaborations, for example, financial savings, access to foreign know-how or access to new foreign market and its customers. Identify potential benefits for specific organizational context and therefore reasons for doing collaborations.

#### 4.1.2 Collaboration model

**Objectives:** The practice of defining a Collaboration model for global collaborations aims at identifying answers for such questions as how to perform global collaboration, according to what scenario and where to. The identified answers aim at matching to specific organizational context and goals. Defining Collaboration model is the starting decision point for companies intending to form global collaborations in the software development domain.

**Potential results and experiences:** Based on the literature review and inputs from the industrial partner three scenarios of launching global collaborations were identified and named Offshore outsourcing, Offshore insourcing and Innovative offshoring [SWG12, PAD07, MSH12, HPT06, VBK14]. The potential results and experiences for each scenario are described below. The following Figures 5 and 6 summarize the description of identified scenarios.

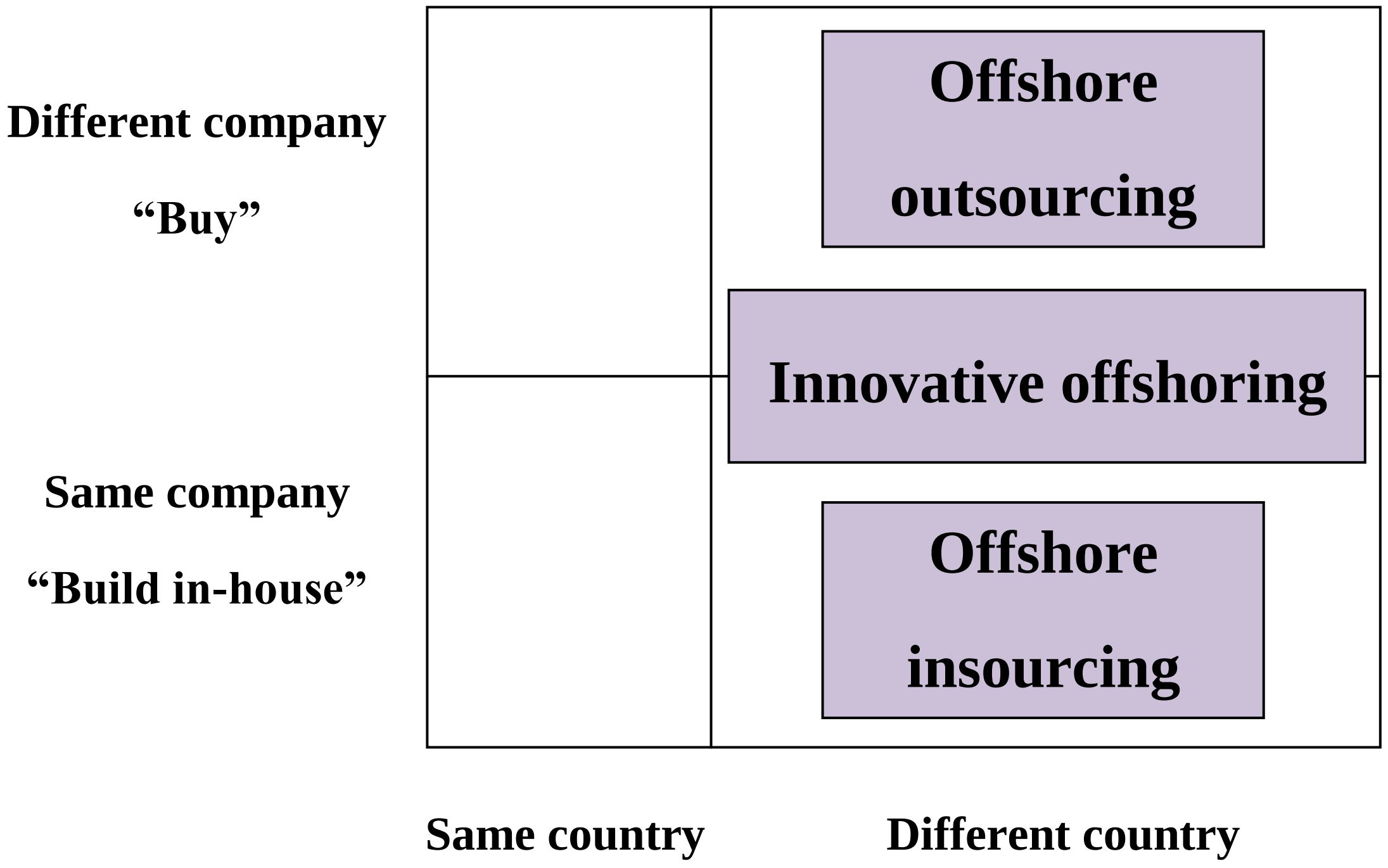

Figure 5. Collaboration scenarios (based on [SWG12, PAD07])

| Scenario 1. Offshore outsourcing | Scenario 2. Offshore insourcing | Scenario 3. Innovative offshoring |
|---|---|---|
| ▪ consuming of resources and development services from an external 3rd party that is located in a foreign country<br>▪ client-subcontractor relationships | ▪ consuming of internal organizational resources that are located in a foreign country<br>▪ product customization<br>▪ responsibility for dedicated market | ▪ consuming of innovative R&D services from the offshore partner company that is located in a foreign country<br>▪ headquarter's product improvement |

Figure 6. Description of the collaboration scenarios

***Scenario 1.***

***Objectives:*** The Offshore outsourcing model refers to consuming of resources and development services from an external 3rd party that is located in a foreign country [SWG12]. This model often represents client-subcontractor relationships between headquarter onsite organization and offshore partner.

***Potential results and experiences:*** Promising benefits of such a collaboration scenario are the reduction of development costs via transferring work to destinations with lower resource costs, the speed-up of development process via performing different development process stages in a parallel way at different locations, the access to worldwide skills and expertise. However, this collaboration scenario requires certain investments for the training of offshore development teams, including knowledge sharing and in particular business domain knowledge transfer. Based on a case study of Prikladnicki et al. [PAD07] conducted at five companies located in Brazil, Canada and the United States operating as a computer company, IT service providers, a provider of enterprise energy management solutions, and a retailing company it was identified that often strategically important activities such as requirements creation and system design in the offshore outsourcing collaboration scenario are kept onsite. Thus the requirements clarification might become troublesome. Additionally, as development process stages activities are globally distributed in a parallel way, the efforts on system integration might be high in the end. The core of this model is based on client-subcontractor relationships between sites, thereby the social ties between collaboration sites are not that important and can stay relatively weak without much investment.

***Scenario 2.***

***Objectives:*** The Offshore insourcing model refers to consuming of internal organizational resources that are located in a foreign country [SWG12]. The company establishes, for instance, a foreign branch in a different country, in order to customize products for a dedicated foreign market. This model is built as peer-to-peer partnership relationships between collaboration sites meaning that know-how and responsibilities are shared between sites [LeM07].

***Potential results and experiences:*** This collaboration scenario gives an opportunity for organizations to access new foreign technologies and enter new markets that might lead to new customers and bigger revenues. However, this model essentially requires investments for building strong social ties and trust between collaboration partners. Case study of Moe et al. [MSH12] at three Scandinavian companies operating in business and engineering domains on collaborations with India and China showed that, in the end, investing into the social ties building might lead to the building of an easier common understanding, corporate spirit and to the increasing of motivation for collaboration at both the onsite and the offshore locations. Also, the offshore insourcing collaboration model requires efforts into mutual knowledge sharing and building equal compatible infrastructure [VBK14]. Case study of Kobitzsch et al. [KRF01] at Tenovis GmbH & Co.KG operating in communication solutions domain on collaboration between Germany and India claims that knowledge transfer should be performed as a continuous activity in global collaborations.

***Scenario 3.***

***Objectives:*** The Innovative offshoring model refers to the consuming of innovative R&D services from the offshore partner company that is situated abroad.

***Potential results and experiences:*** The model gives an access to the innovative foreign know-how capabilities that might be not obtainable onsite [FBD08]. These innovative services could, for instance, aim at improving or optimizing the headquarter’s product. However, this model requires efforts and investments into the knowledge transfer as well as building social ties and trust between collaboration partners. As a critical point of this model, the end system integration can be considered quite troublesome, especially when the end system combines a few innovative product parts and they were developed earlier on at different locations.

**Actions:** Identify and choose what collaboration scenario might be more beneficial for the specific organizational context. Such as, choose one of the three most typical scenarios - Offshore outsourcing, Offshore insourcing or Innovative offshoring. Customize existing scenarios for specific organizational goals if needed.

#### 4.1.3 Foreign legal system

**Objectives:** The practice of investigating a foreign legal system aims at making organizations intending on setting-up global collaborations familiar with legal matters of collaborations, particularly regarding the contract and the Intellectual Property (IP) laws.

**Potential results and experiences:** The knowledge of legal matters aims at helping to avoid possible problems with data confidentiality and copyright protections in situations of knowledge and know-how transfer. Furthermore, this is an essential high level step when organizations intend to establish new foreign branches of the company [VBK14].

**Actions:** Investigate the foreign legal system in the desired location(s) of global collaboration, especially concerning IP and contract law.

#### 4.1.4 Vendors

**Objectives:** The practice of selecting suitable vendors for doing global collaborations aims at helping organizations to choose right collaboration partners with desired expertise, efficient capabilities and sufficient technological infrastructure [SMS14].

**Potential results and experiences:** The investigation of vendors' capabilities and infrastructure resources early on aims at helping to avoid potential problems that might occur later on when software development process is already ongoing [VBK14]. Moreover, insufficient vendors' expertise might make an influence of needed investments into software development projects and become a reason for hidden projects costs, that's why it is recommended to "learn" your partners early on.

**Actions:** Choose appropriate vendor(s) with sufficient infrastructure, capabilities and expertise needed for the chosen collaboration model.

#### 4.1.5 Budget plan

**Objectives:** The practice of planning the financial budget for global collaborations aims at making organizations aware of the collaboration costs including possible risks and therefore also hidden costs [VBK14]. One of the main drives for performing GSD collaborations by companies could be financial savings, therefore financial planning should be considered early on [SMS14].

**Potential results and experiences:** Financial planning, especially the analysis of hidden GSD collaboration costs, might help organizations to figure out if a global collaboration promises potential cost savings or not. Those investigations might lead to high level decisions if it is beneficial for the company to set-up global collaborations in the domain software development or if it is more profitable to stay onsite and do traditional co-located development.

**Actions:** Define a budget plan for doing global software development projects. Include possible hidden costs such as communication tools, face-to-face visits.

## 4.2 Collaboration structure

The aspect *Collaboration structure* is aimed at determining the approach of development task allocation between locations based on collaboration goals, at creating roles and responsibilities along with the way of distributing them; at defining an organizational structure and peer-to-peer connections between collaboration sites [SMS14, Smi06, RCM12, Nis04, FQA07, CuP06, LMR08].

This aspect is aimed at creating a clear understanding of the division of work, roles and responsibilities between collaboration sites. Therefore, it might help to decrease coordination and project management efforts when the actual software development process is ongoing [SMS14]. It is important to consider how to organize collaboration structures early on, ideally already at the planning stage of setting up global collaborations [SMS14].

With the aspect of *Collaboration structure* two main practices were extracted from the literature – to define the approach to distributed process breakdown and task allocation, so called General task distribution, and to determine and specify the Organizational structure and peer-to-peer links between sites [SMS14].

### 4.2.1 General task distribution

**Objectives:** The practice of defining a General task distribution model for global collaborations aims at identifying a way of development process breakdown with following task allocations between collaboration sites. Moreover, it aims at creating and dividing roles and responsibilities between locations based on collaboration goals and existing expertise and resources.

**Potential results and experiences:** Based on Šmite's case study in a Latvian software company [Smi06, Smi05], Nissen's case study report from an inter-organizational cooperation in telecommunications domain between a German company and an Indian IT-service provider [Nis04], and the research of Faiz et al. [FQA07], three models of process-based task distribution between collaboration sites were identified [SMS14]. These three models are suitable for the collaboration scenarios described earlier in the practice *Collaboration model* for the aspect

*Strategy* (see section 4.1.2) [Smi06, Nis04, FQA07, CuP06]. The potential results and experiences for each scenario of task distribution are described below.

***Scenario 1.***

***Objectives:*** The first scenario of general task distribution, the so-called Offshore outsourcing model (see section 4.1.2), refers to the task division model where most of the intellectual work stays mainly at the onsite location and only actual software development engineering tasks are fully transferred to the offshore location [Smi06, Nis04, FQA07, CuP06, SMS14]. Typically, in this model system requirements creation is fully performed onsite. System analysis and design activities are often performed cooperatively with an offshore partner. System implementation activities such as coding and system testing might be fully transferred to an offshore location or done jointly, for instance, divided between sites by system modules or subsystems (Figure 7).

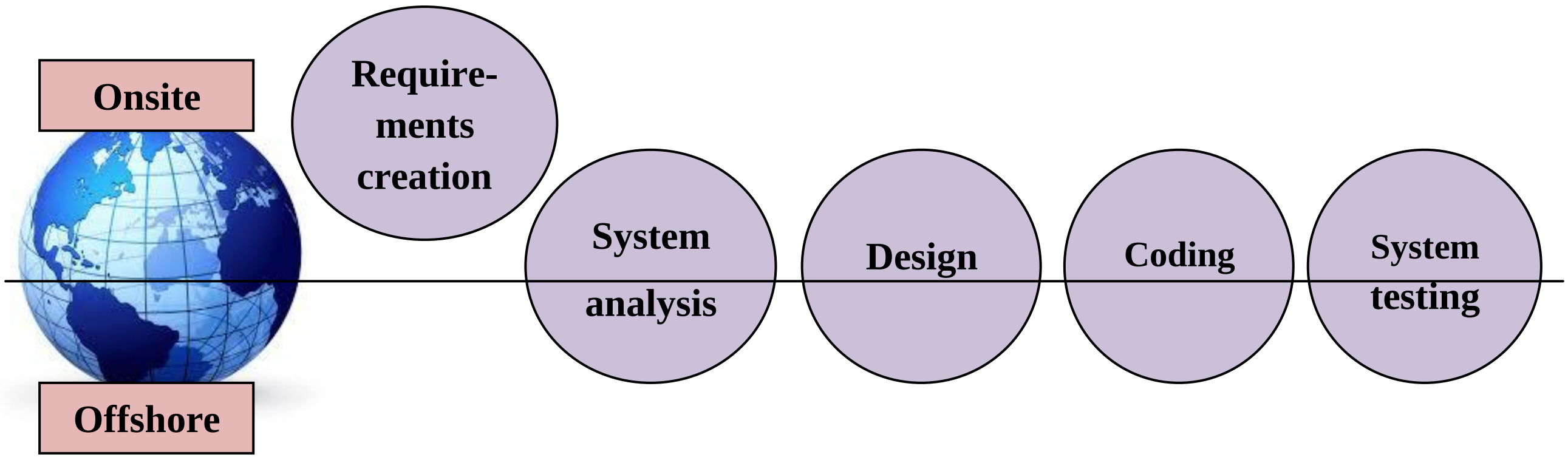

Figure 7. General task distribution: Offshore outsourcing scenario (based on [Smi06, CuP06])

***Potential results and experiences:*** Promising benefits of such a task distribution scenario are the reduction of development costs and the speed-up of development process through the access to labor pool at the offshore destinations. Moreover, as a benefit of this scenario, strategical know-how is kept onsite. The main challenges of this kind of task distribution consist of troublesome requirements clarification between collaboration sites, effortful system integration and bug fixing [SMS14]. Also, coordination and control efforts in this scenario might be high [SMS14].

***Scenario 2.***

***Objectives:*** The second scenario of general task distribution, the so-called Offshore insourcing model (see section 4.1.2), refers to the task division model where, in contrast to the first scenario, both intellectual and implementation work are performed as joint activities between collaboration partners [Smi06, Nis04, FQA07, CuP06, SMS14]. In this model, typically, system requirements creation, system analysis and design activities are performed jointly, while system implementation activities might be fully transferred to the offshore destinations or done by cooperation (Figure 8). Depending on the work allocation within the company the task distribution between sites can be very similar to the first scenario.

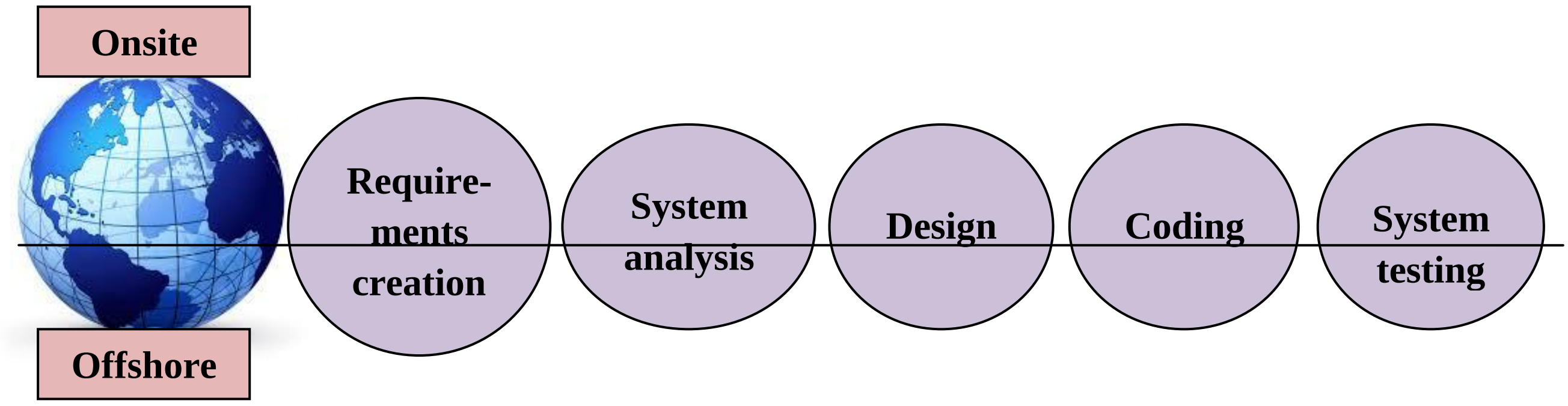

Figure 8. General task distribution: Offshore insourcing scenario (based on [Smi06])

***Potential results and experiences:*** Such a task distribution scenario might suit the product customization and localization goals [SMS14]. This scenario helps to create a partnership type collaboration between sites and helps to achieve better common understanding and the building of social ties between locations [SMS14]. However, this model requires good domain business knowledge from the offsite location that might be effortful to create [SMS14]. In addition, a lot of effort should be put into communication, trust and social relationships building.

***Scenario 3.***

***Objectives:*** The third scenario of general task distribution, the so-called Innovative offshoring model (see section 4.1.2), refers to the task division model where the offshore partner performs most of the intellectual and implementation work such as R&D, requirements creation, system analysis and design, and actual implementation tasks (Figure 9) [Smi06, Nis04, FQA07, CuP06, SMS14].

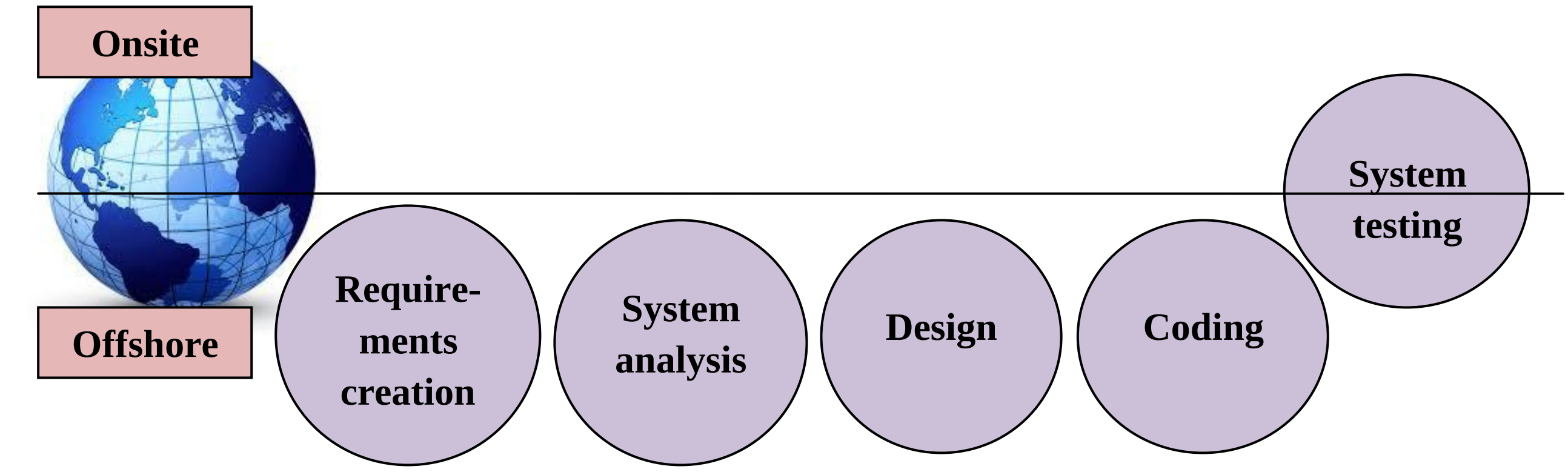

Figure 9. General task distribution: Innovative offshoring scenario (based on [Smi06])

***Potential results and experiences:*** This model might be based on prototyping strategy that refers to the creation of innovative prototypes by the offshore site [SMS14]. Later the developed prototype can be used onsite as a base for building new functionality or a complete product on top [SMS14]. Therefore, this model might suit product improvement and optimization purposes. The challenges of this model consist of the need for the good domain business knowledge from the offshore partner site that requires investments and efforts. Furthermore, trust and social ties building are essential for this model in order to create collaborative relationships between partners.

**Actions:** Based on the collaboration model and its goals, identify and choose what way of task distribution might be more beneficial for the specific organizational context and available resources. Based on the chosen model of task allocation, identify and divide roles and responsibilities between collaboration sites.

### 4.2.2 Organizational structure and peer-to-peer links

**Objectives:** The practice of specifying the Organizational structure and peer-to-peer links aims at creating and assigning in detail roles, responsibilities and communication channels between collaboration sites at all organizational levels – management, projects and team levels (Figure 10) [PaL03, SMS14].

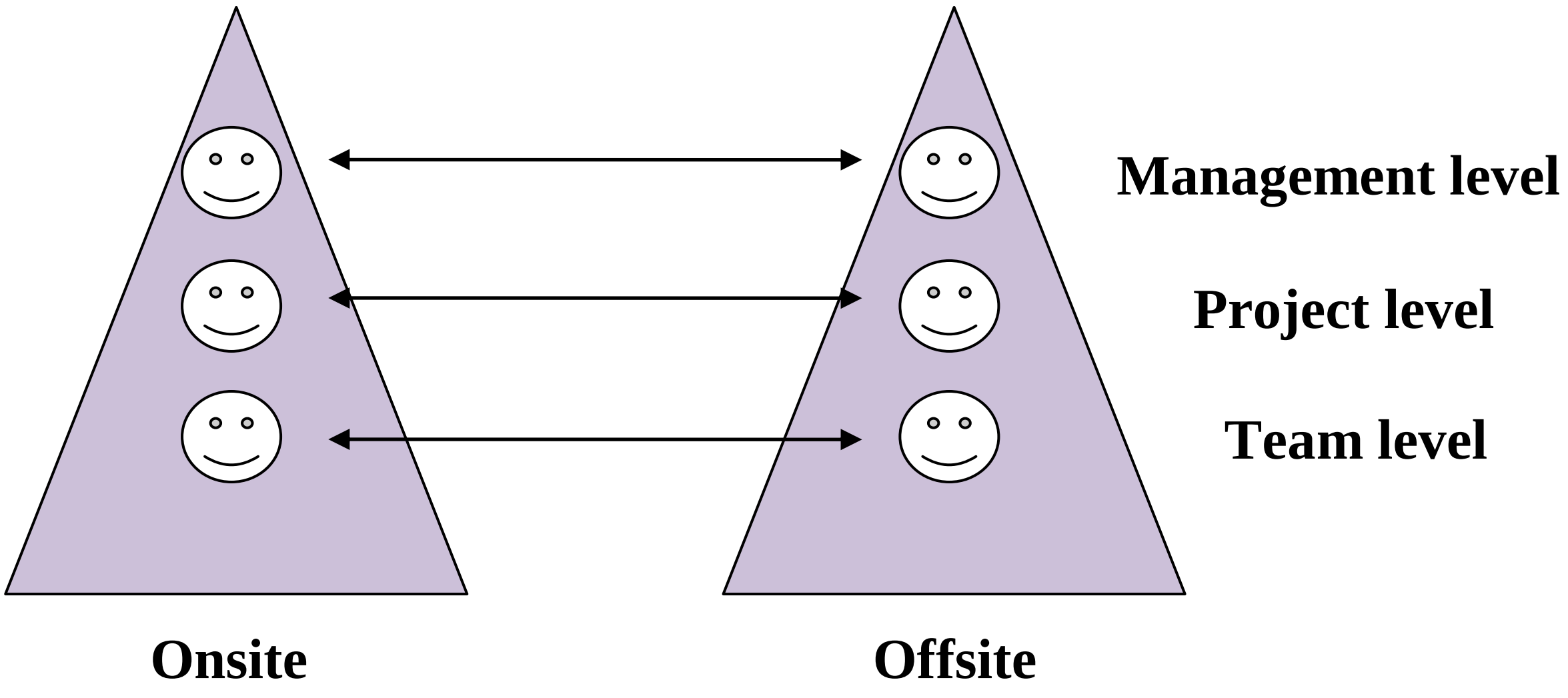

Figure 10. Peer-to-peer links between collaboration sites at all organizational levels [PaL03]

**Potential results and experiences:** A defined and documented organizational structure including peer-to-peer links between collaboration sites aims at easing coordination and control efforts, making information flow between sites more transparent and traceable [PaL03, CuP06,

FQA07, Bra07, SMS14]. Peer-to-peer links aim at stimulating communication between locations and making it more frequent, faster and clearer [PaL03]. Therefore it helps to achieve common understanding between sites faster. This could eventually lead to saving a significant amount of effort and investments needed for setting up global collaborations [SMS14].

**Actions:** Based on the identified earlier collaboration and task distribution models, create, assign and document detailed roles and responsibilities between collaboration sites. Then specify communication channels, interface points and type of information needed to be exchanged between all identified roles. Based on the research of Faiz et al. [FQA07] and the case study of Braun on the IT project at T-Systems/Deutsche Telekom [Bra07] two alternative models of organizational structure in global collaborations were created. The first model of organizational structure is designed for the Offshore outsourcing scenario (see section 4.2.1 – Scenario 1), and the second model is suitable for the Offshore insourcing and Innovative offshoring scenarios (see section 4.2.1 – Scenarios 2 and 3). The list of roles, responsibilities and peer-to-peer connections between collaboration sites are summarized in the following Figure 11, Figure 12 and Figure 13.

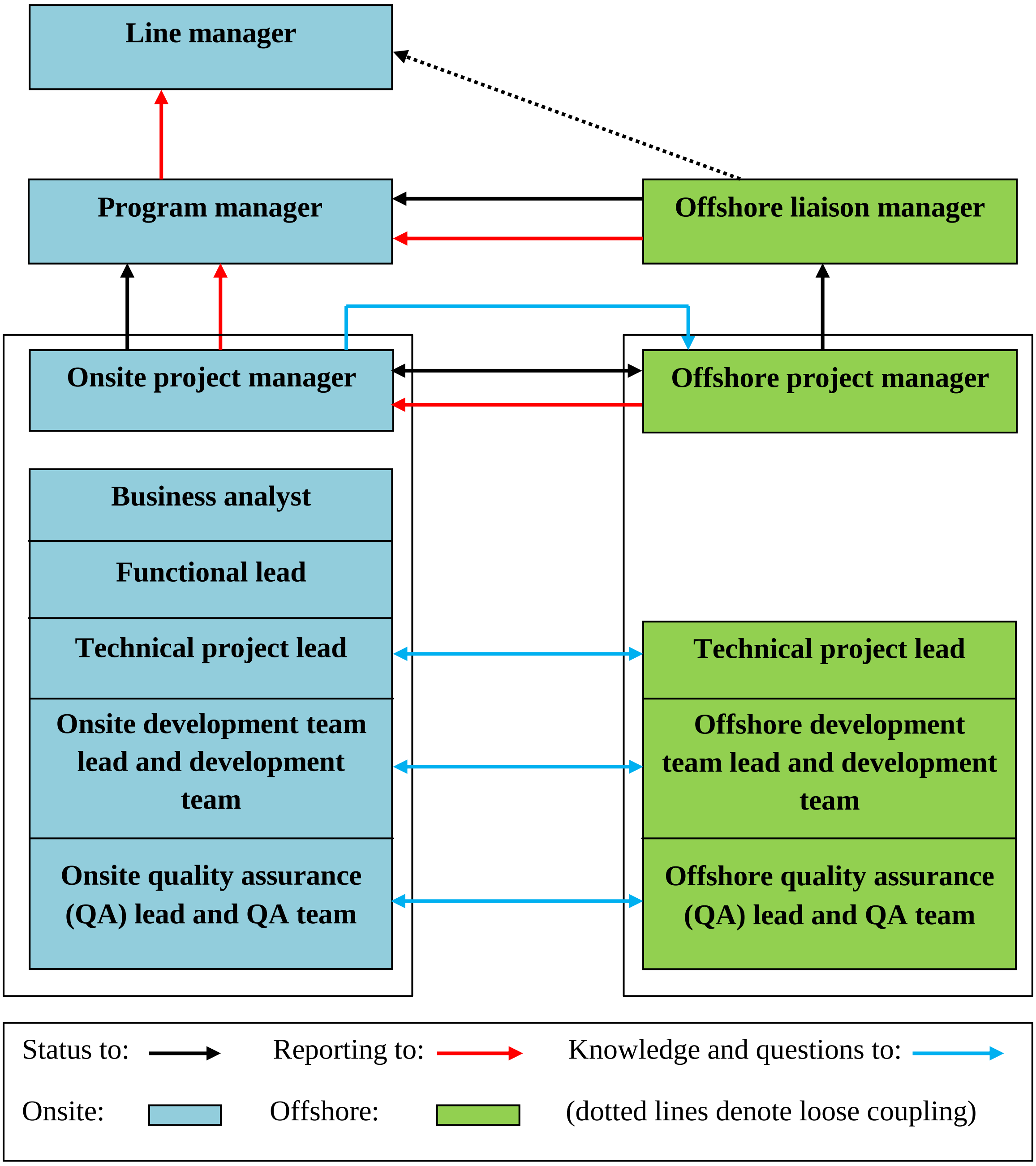

Figure 11. Organizational structure: Offshore outsourcing scenario (based on [Bra07, FQA07])

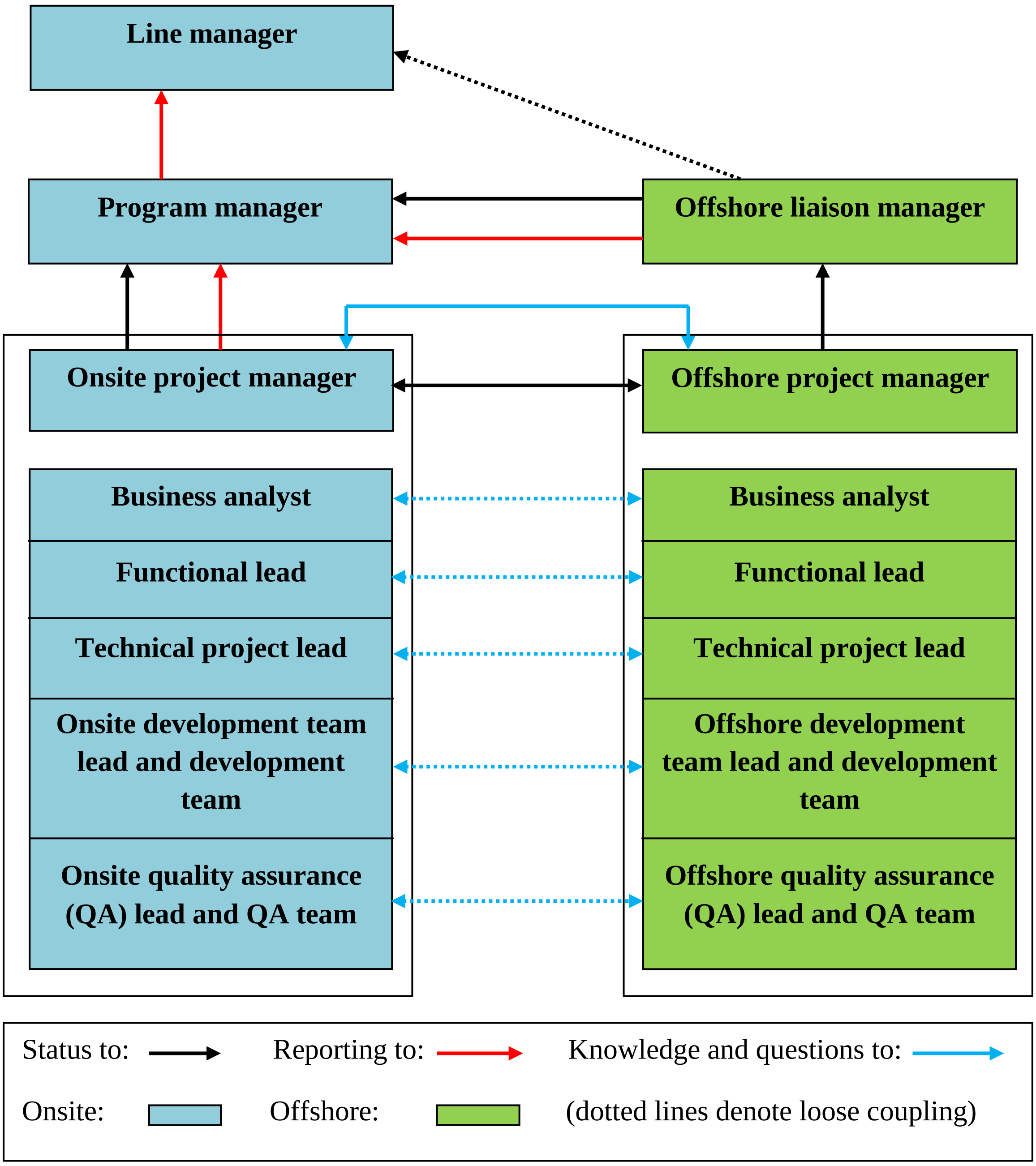

Figure 12. Organizational structure: Offshore insourcing
and Innovative offshoring scenarios (based on [Bra07, FQA07])

| Role | Responsibilities |
|---|---|
| **Management level** | |
| **Line manager** | definition of product line strategy, product line delivery and offshoring strategy |
| **Program manager** | overall program execution, definition of individual projects based on planned product line, offshore governance, budgeting and scheduling |
| **Offshore liaison manager** | offshoring goals and objectives at program level, management of offshore subcontractor(s) |
| **Project level** | |
| **Project manager** | owner of the project assignment, execution of the whole project |
| **Team level** | |
| **Business analyst** | definition of Business Concept, System Analysis |
| **Functional lead** | System Analysis, Functional Design, creation of functional specifications for the project |
| **Technical project lead** | detailed system design based on functional specifications |
| **Development team lead** | execution of work packages, management of the development team, reporting, work tracking |
| **Development team** | software system's development |
| **Quality assurance lead** | execution of testing and quality assurance of work packages, management of the quality assurance team |
| **Quality assurance team** | software system's testing and quality assurance |

Figure 13. List of roles and responsibilities in global collaborations (based on [Bra07, FQA07])

## 4.3 Product structure

The aspect *Product structure* is aimed at determining how the product architecture could be adapted for global software development compared to co-located development, the product ownership boundaries between collaboration sites and how modifications to the product at one location might affect work at other locations [VBK14, RCM12, Smi05, SMS14].

This aspect of defining product architecture and product ownership has an impact on division of work and responsibilities between collaboration sites by making them clearer. Therefore, it is expected to improve the communication and creation of mutual understanding between sites, reduce coordination efforts and help to avoid rework and duplications [SMS14, VBK14, RCM12, Smi05, Sal09, LMR08]. Defining the product structure is an important step for organizations and needs to be considered early on, already at the planning stage of global collaborations [SMS14]. The aspect is tightly connected with the *Collaboration structure* [SMS14].

With the aspect of *Product structure* two main practices were extracted from literature – to define the approach to Product architecture and product ownership between locations and to specify Product-based work distribution between development teams at different collaboration locations [SMS14].

### 4.3.1 Product ownership and architecture

**Objectives:** The practice of defining Product ownership and architecture for global collaborations aims at identifying how the product structure and architecture should be adapted for global collaborations and the product ownership boundaries between collaboration sites.

**Potential results and experiences:** The Product ownership and architecture model has a strong influence from which collaboration model is chosen by organization and how the development process breakdown and general task distribution are organized. Moreover, the Product ownership and architecture model itself has a great influence on system architecture and detailed work distribution between sites that later on often determines the communication, coordination, system integration efforts. The following Figure 14 summarizes which aspects are connected with the Product ownership and architecture practice.

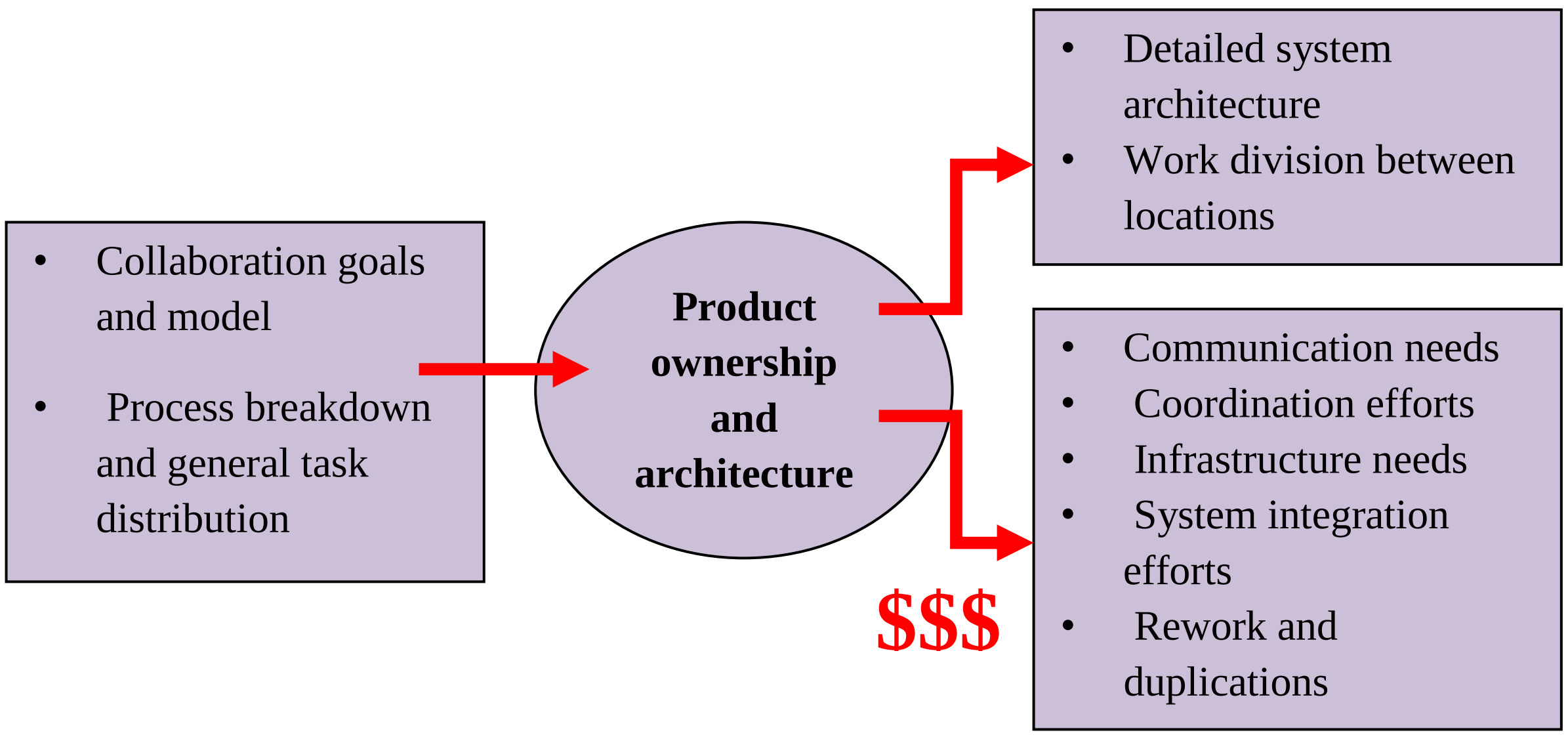

Figure 14. Connection of the Product ownership and architecture with other aspects of global collaborations (based on [Gor14])

Based on the literature review and inputs from the industrial partner three scenarios of organizing Product ownership and architecture were identified [SPF12, HeM03, PAP06, Smi05, DIS07, SWG10, VBK14, OKW08, LeM07, MoW01, RCM12, HoM07, HPT06, AFH05, LLA07]. These three models are suitable for the collaboration scenarios described earlier in the practice *Collaboration model* for the aspect *Strategy* (see section 4.1.2). The potential results and experiences for each scenario are described below.

***Scenario 1.***

***Objectives:*** The first scenario of product ownership and architecture, that is suitable for the Offshore outsourcing model (see section 4.1.2), refers to the model where one core product with full onsite product ownership is developed [SMS14]. The offshore partner is responsible for some engineering tasks in the product development lifecycle (Figure 15) [LeM07, MoW01, HPT06]. Modular system architecture in this model is expected to ease the work transfer to offshore destinations.

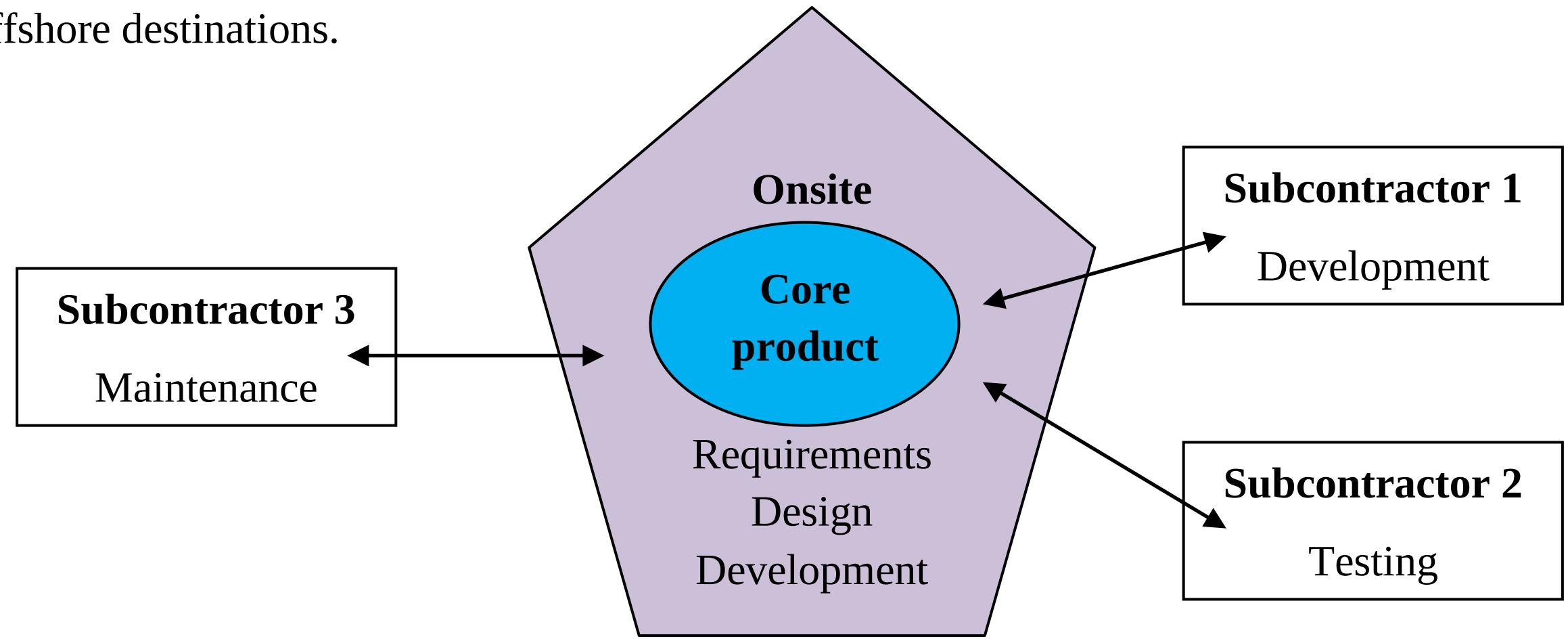

Figure 15. Product ownership and architecture: Offshore outsourcing scenario (based on [Gor14])

***Potential results and experiences:*** Promising benefits of such product ownership and architecture scenario are the speed-up of development process through performing some engineering tasks in the product lifecycle in a parallel way that also helps to achieve less work dependencies and more freedom between collaboration sites. A case study by Leszak et al. on embedded product development in telecommunications domain in Alcatel-Lucent between Germany and China [LeM07] reports as a benefit of such scenario that strategically important know-how is kept onsite. The reported main challenges of this kind of product ownership and architecture model consist of high coordination and control efforts, effortful system integration and needed quality inspections for the tasks performed at different offshore destinations [LeM07, MoW01, HPT06].

***Scenario 2.***

***Objectives:*** The second scenario of product ownership and architecture, that is suitable for the Offshore insourcing model (see section 4.1.2), refers to the model of product line architecture with variants which are built on top of a core product [SMS14]. The core product is developed onsite, while the product variants are transferred for the full development to the offshore destinations. The product ownership of the core product might be kept onsite, while the offshore partner owns and is fully responsible for the allocated variant from the product line (Figure 16) [BBA13, SMS14]. Such product variant can be customized and market-specific [SMS14]. In this model both the onsite and offshore parties have a high responsibility for the product, and thereby global collaboration might have a form of peer-to-peer partnership [Smi05, LeM07, SMS14].

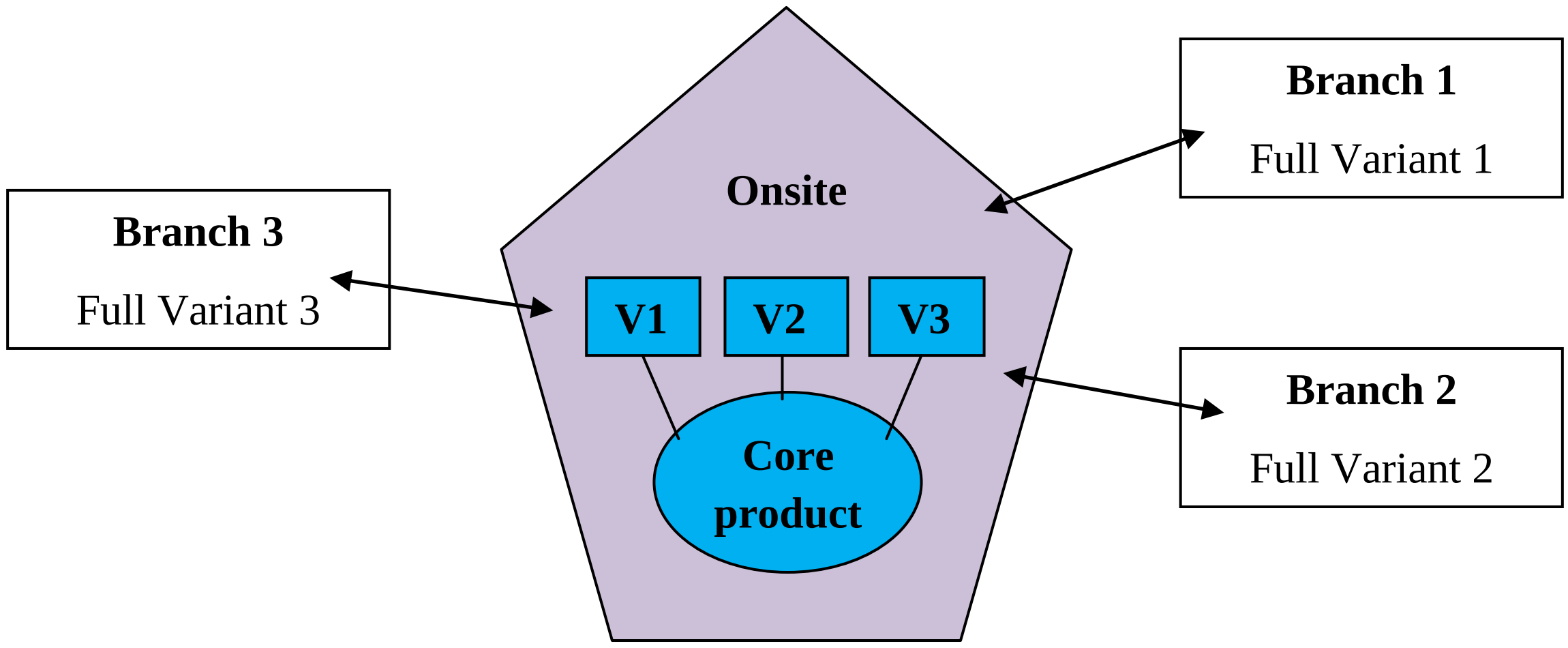

Figure 16. Product ownership and architecture: Offshore insourcing scenario (based on [Gor14])

***Potential results and experiences:*** Such a product ownership and architecture scenario might suit product customization and localization goals. Based on the case studies of Lings et al. at eight companies within Europe operating in the financial domain, car manufacturing, IT-manufacturing and telecommunications domain [LLA07], the research of Mockus et al. at Lucent Technologies [MoW01] and the study of Bhadauria et al. on successful German-Indian collaboration at Siemens [BBA13] it was identified that such a scenario helps to achieve less work dependencies between collaboration sites, to get more flexibility, to reduce coordination and communication efforts. However, this model requires good domain business knowledge at offshore destinations that might be effortful to create. Moreover, a lot of effort should be put into the building of trust and social ties. As a "critical point" in this scenario the offshore destinations being responsible for the whole product variant might get more and more knowledge and power and become totally independent. Thereby, in order to avoid this, the situation should be monitored and controlled by the onsite organization.

***Scenario 3.***

***Objectives:*** The third scenario of product ownership and architecture that is suitable for the Innovative offshoring model (see section 4.1.2), refers to the development of new innovative product(s) or prototype(s) by the offshore site that later on is used onsite for building products with different functionality on top of the innovative prototype(s) (Figure 17) [SMS14]. The product ownership boundaries and responsibilities between sites might differ depending on the initial collaboration goals and the chosen collaboration model [SMS14].

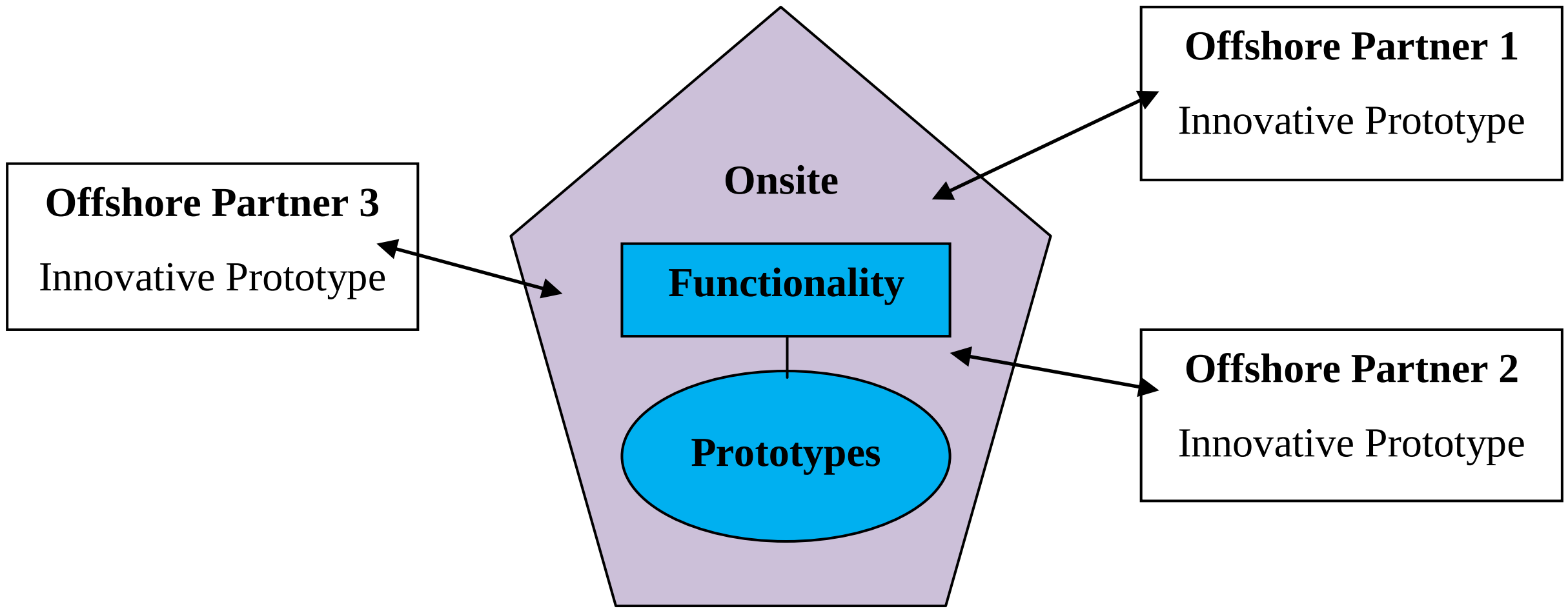

Figure 17. Product ownership and architecture: Innovative offshoring scenario (based on [Gor14])

***Potential results and experiences:*** This product ownership and architecture scenario promises access to foreign know-how and therefore might suit product optimization and improvement goals. Such a scenario is expected to reduce coordination, communication efforts and to give more flexibility to collaboration sites. The challenges of this scenario consist of building trust and social ties that are essential in order to create collaborative relationships between partners. A "critical point" in this scenario could be that the offshore partners being quite independent might get more and more power and, thereby, might collude. Therefore, in order to avoid the situation of collusion, the onsite organizations should monitor their offshore partners and create peer-to-peer partnership-based relationships.

**Actions:** Based on the development process breakdown and following task distribution, identify and choose what model for product ownership and architecture might be more beneficial for the specific organizational context and collaboration goals. Such as, choose one of the three most typical scenarios presented above or define a combination of the scenarios that fits organizational context and goals better. Customize existing scenarios for specific organizational goals if needed.

### 4.3.2 Product-based work distribution

**Objectives:** The practice of defining Product-based work distribution between GSD teams at collaboration sites aims at identifying the approach to the distribution of software development work between locations based on detailed software system architecture.

**Potential results and experiences:** Detailed work distribution aims at achieving rational work distribution between locations based on available resources and expertise. The main suggestions are to use modular architecture and decoupling principles that allow constructing well-defined software work packages which can be easily distributed between different locations [SMS14]. Salger [Sal09] demonstrates an example structure of the software work package based on the experience at Capgemini sd&m [SMS14]. The software work package might consist of the following parts: *Software requirement specifications* describing use cases, user interface, domain objects, and specifications of functional test cases; *Design artefacts* including an external technical view on a software module and internal high level design view; *Project management artefacts* containing a list of work units described in earlier parts, schedules, budget, definition of quality objectives and work acceptance criteria [Sal09, SMS14]. Such a well-defined work package structure promises to ease work transfer between locations, to achieve low dependencies between sites during actual implementation work, to avoid rework and duplications [SMS14, VBK14, SWG10, AFH05, LLA07, RCM12, Smi05, LeM07, SPF12, ChE07, NDC13, HoM07,

MoW01]. This way it might be possible to reduce communication and coordination needs between different locations [SMS14]. However, system integration could become a troublesome bottleneck [SMS14].

**Actions:** Based on the model for development process breakdown and general task distribution, the approach to product ownership and architecture, identify the way of detailed work distribution between collaboration sites and create software work packages for transfer to different locations based on available resources and expertise.

## 4.4 Coordination

The aspect *Coordination* is aimed at integrating software development work in a way that each involved unit contributes to the completion of the overall task [SMS14]. Coordination procedures describe how collaboration sites communicate between each other in order to complete commonly defined tasks and to achieve collaboration goals [AFH05, HoM07, SMS14].

The *Coordination* aspect represents the set of activities that aim at managing dependencies within the global software development project workflow, so the work can be completed effectively and be inside a financial and technological budget [SMS14]. According to a study by Nguyen-Duc and Cruzes [NDC13] on the software development collaboration between Norway and the USA such dependencies in a GSD context might include technical views such as system integration, configuration change management; temporal issues such as synchronization of schedules, deliveries between sites; software development process organization; resource distribution such as infrastructure, budget, or development tools [SMS14]. This aspect promises to be very important for setting up global collaborations, because coordination and project management efforts might become a cause of project hidden costs [SMS14]. Therefore coordination activities need to be carefully planned and performed by the organizations on an everyday basis starting from a planning phase [SMS14]. For instance, differences in organizational policies, lack of common processes, variation of coding and testing standards between collaboration sites might affect coordination and project management efforts, and also might lead to insufficient end product quality and additional costs [NDC13, SMS14].

Within the aspect of *Coordination* two main practices were extracted from the literature – to define the approach to Project management and to specify Project control procedures [SMS14].

### 4.4.1 Project management

**Objectives:** The practice of defining Project management approach aims at planning and organizing software development project-related activities in such a way that they lead to successful work completion [SMS14]. Such activities might include creating shared synchronized understanding of main milestones between collaboration sites, concrete tasks to perform, deliveries schedules, project budget constraints, peer-to-peer contact links between collaboration sites, and managing the whole project execution [NDC13, SMS14].

**Potential results and experiences:** Project management aims at being a mechanism that integrates software, human, and economic relations in order to use existing technology, resources, time, capabilities in the most productive and effective way [Cas10, SMS14]. Defined approach to Project management helps to ease management efforts inside software projects, reduce communication efforts, increase project visibility, avoid possible misunderstandings between collaboration sites during the software development work and build up social ties between sites. Therefore the clearly defined way of performing Project management in GSD projects might lead to costs savings when setting up global collaborations.

**Actions:** Based on the collaboration model, the identified task distribution and the product architecture identify and customize the approach to Project management activities. Based on a case study by Hossain et al. in an Australian-Malaysian cooperation on software product development [HBV09] two alternative possible ways of performing coordination processes were identified [SMS14].

The first Project management model is suitable for the Offshore outsourcing scenario (see section 4.2.1 – Scenario 1). The model refers to a high degree of defined standardization policies, including, for instance, documented standard instructions to software development process, organizational structure, coding, documentation, testing assurance, and change management system [VBK14, BBA13, AFH05, PAD07, Smi05, CuP06, NDC13, Cas10]. This model also requires direct supervision with tasks and instructions and centralized project organization for the remote offshore teams from the headquarter organization that means the minimum of freedom for the offshore partners. This model aims to avoid misunderstanding at work, reduce delays and avoid task conflicts between sites. However, the standardization is hard to achieve and requires time and investments. As a critical point, this model creates “we” and “you” relationships between collaboration sites that make trust building harder and more effortful.

The second Project management model is suitable for the Offshore insourcing and the Innovative offshoring scenarios (see section 4.2.1 – Scenarios 2 and 3). The model refers to a high degree of mutual adjustment when collaboration is based on building trust and social ties between partners.

The model uses informal communication, jointly performed software project activities and decisions, frequent face-to-face visits and staff exchanges, knowledge transfer and decentralized project organization. This approach aims at creating the "one team" mindset between collaboration partners. However, the critical and effortful sides of the model are the achieving project transparency, adapting to the new team and work culture, challenges in creating clear mutual understanding [HBV09, LeM07].

### 4.4.2 Project control

**Objectives:** The practice of defining Project control procedures aims at identifying the process of monitoring work status and ensuring that the work process goes in the right direction according to the planned budget, timeframes and quality expectations [AFH05, SMS14].

**Potential results and experiences:** Project control procedures aim at defining a formal reporting structure concerning updates, changes and escalation path in the GSD projects that help to achieve the visibility of software project status and monitor work progress, detect project bottlenecks and work conflicts situations early on and react on them [SMS14, VBK14, Bha13, RCM12, NDC13]. The practice is tightly connected with the Project management approach.

**Actions:** Based on the collaboration model, task allocation and identified project management approach choose what type of control procedures might be suitable for particular collaboration goals. For example, use standardization mechanisms with strict control from the headquarters' side or jointly performed mutual adjustment methods.

## 4.5 Development process

**Objectives:** The aspect and the practice itself of defining a *Development process* model are aimed at defining and/or customizing the model for software development activities between the collaboration sites [SMS14].

**Potential results and experiences:** Defining a software development process model aims at clarifying roles and responsibilities, the level of independency between sites, and the product quality expectations [SMS14]. Frictions such as role confusions can be avoided [SMS14]. Furthermore, a development process can affect coordination and communication efforts, infrastructure needs, change management mechanisms and system integration efforts [SMS14].

Based on the literature review and inputs from the industrial partner three scenarios of organizing a Development process model were identified. These three models are suitable for the collaboration and task distribution scenarios described earlier (see section 4.2.1). The potential results and experiences for each scenario are described below.

***Scenario 1.***

***Objectives:*** The first scenario of a software development process model that is suitable for the Offshore outsourcing model (see section 4.2.1 – Scenario 1), refers to the model where collaboration sites might keep their own development processes if they are already established and well-working [RCM12, PaL03, SMS14]. It is recommended to mainly define the development processes at the interfaces between the collaborating sites and not to aim at unification of all processes at all sites, especially when the sites belong to different organizations [SMS14]. A software development process model for the first scenario is illustrated in the Figure 18.

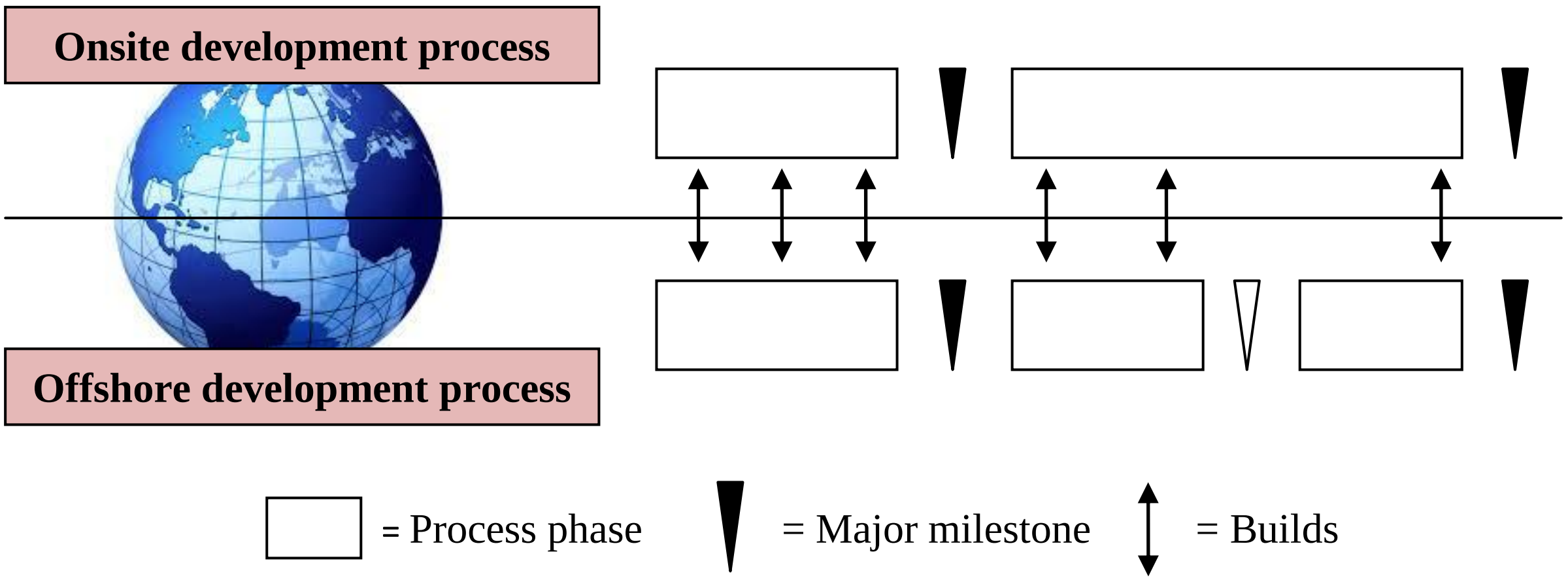

Figure 18. Software development process model: Offshore outsourcing scenario (based on [PaL03])

***Potential results and experiences:*** Such a model might save infrastructure costs and offer flexibility to the collaboration sites. However, based on the interview studies of Paasivaara and Lassenius [PaL03] at six Finnish companies, it is recommended to synchronize project milestones and schedules for important product deliveries in order to achieve project transparency and traceability and detect problems early on. Therefore the coordination and control efforts might be high. Moreover, this model suggests using continuous integration and early feedback on the quality of developed software that could reduce the system integration efforts in the end [RCM12, PaL03].

***Scenario 2.***

***Objectives:*** The second scenario of a software development process model, that is suitable for the Offshore insourcing model (see section 4.2.1 – Scenario 2), refers to the model where standardized guidelines for a common software development process and tools between sites are established (Figure 19) [VBK14, SPF12, Cas10, SMS14].

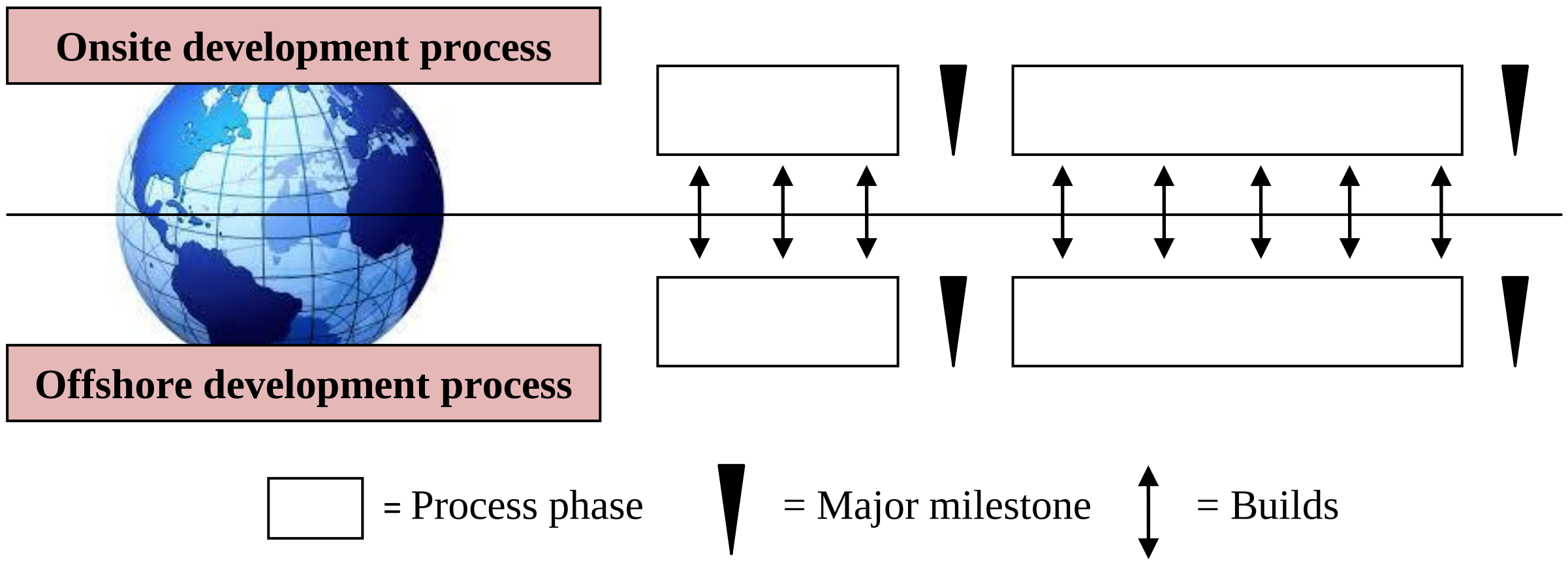

Figure 19. Software development process model: Offshore insourcing scenario (based on [PaL03])

***Potential results and experiences:*** Based on the case study of Prikladnicki et al. [PAD07] at 5 companies located in Brazil, Canada and the United States operating as a computer company, an IT service provider, a provider of enterprise energy management solutions and a retailing company, the study of Höfner and Mani [HoM07] at an offshore software development center in India for a German organization and the study by Nguyen-Duc and Cruzes [NDC13] on the software development collaboration between Norway and the USA, such a development process model promises to create joint corporate standards, unification of all processes and therefore reduce coordination efforts. In addition, standardization promises to achieve better software product quality. However, this model requires investments into infrastructure, creation of standardized processes, and domain knowledge sharing.

***Scenario 3.***

***Objectives:*** The last third scenario of a software development process model that is suitable for the Innovative offshoring model (see section 4.2.1 – Scenario 3), refers to the model where collaboration sites have a high degree of freedom for development process model definition. Therefore collaboration sites might keep their own development processes and only synchronize important milestones and delivery schedules (Figure 20) [RCM12, PaL03, SMS14].

***Potential results and experiences:*** Such a scenario gives a lot of flexibility to the collaboration sites and saves infrastructure costs. However, the coordination and control efforts as well as system integration efforts might be high. Moreover, this model requires sites to make quality inspections in order to monitor the quality level of the developed software [SMS14].

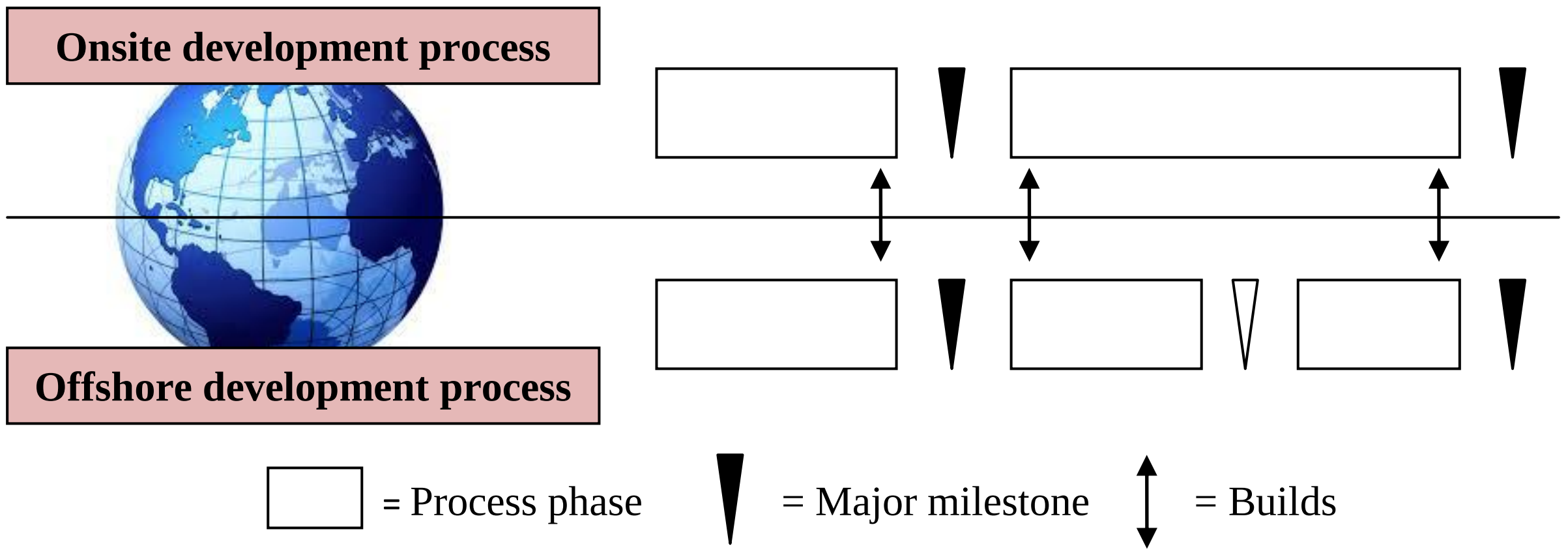

Figure 20. Software development process model: Innovative offshoring scenario (based on [PaL03])

**Actions:** Based on the collaboration model, the task distribution and the product architecture identify and choose what model for software development process might be more suitable and beneficial for the specific organizational context and collaboration goals. For example, choose one of the three most typical scenarios presented above. Customize existing scenarios for specific organizational goals if needed.

## 4.6 Communication

The aspect *Communication* is aimed at addressing all kinds of communication activities between the collaboration sites. Communication can be seen as the exchange of information that helps to reach a common shared understanding between remote sites, including information and knowledge sharing [AFH05, HoM07, SMS14].

Global software development collaborations are to a large degree human-based, so that communication becomes crucial and needs to be considered early on, i.e., starting from the planning and negotiating phases of the collaboration till its full establishment and maintenance [SMS14]. Numerous industrial case studies report the importance of communication in the GSD context and usually come to the conclusion that communication is the number-one problem in global collaborations [SMS14]. Studies focusing on communication include the study by Mettovaara et al. in Nokia and Philips [MSL06], the case study by Leszak and Meier on embedded product development in the telecommunications domain in Alcatel-Lucent between Germany and China [LeM07], the study by Paasivaara and Lassenius based on interviews in 8 global software projects distributed across Europe, North America and Asia [PaL03], and the study by Oshri et al. at LeCroy (Switzerland and USA), SAP (India and Germany), and Baan (India and The Netherlands) [OKW07].

Geographical distances between teams in global collaborations cause difficulties with using traditional communication paths such as face-to-face meetings or informal communication [SMS14]. Therefore the collaboration sites often need to use asynchronous ways of communication (different communication tools such as E-mail, blogs, informal chats) and phone/video conference calls (Figure 21) [SMS14]. Such communication ways that are adjusted to a global context might bring certain risks and challenges like misunderstanding, work delays, unnecessary work, reduced trust, or absence of teamness and partnership feeling [SMS14]. The challenges might result in additional project costs, customer dissatisfaction and difficulties with setting up long-term global collaborations [SMS14]. Thereby it is necessary for organizations to plan communicational strategy early on in the first stages of collaborations [SMS14].

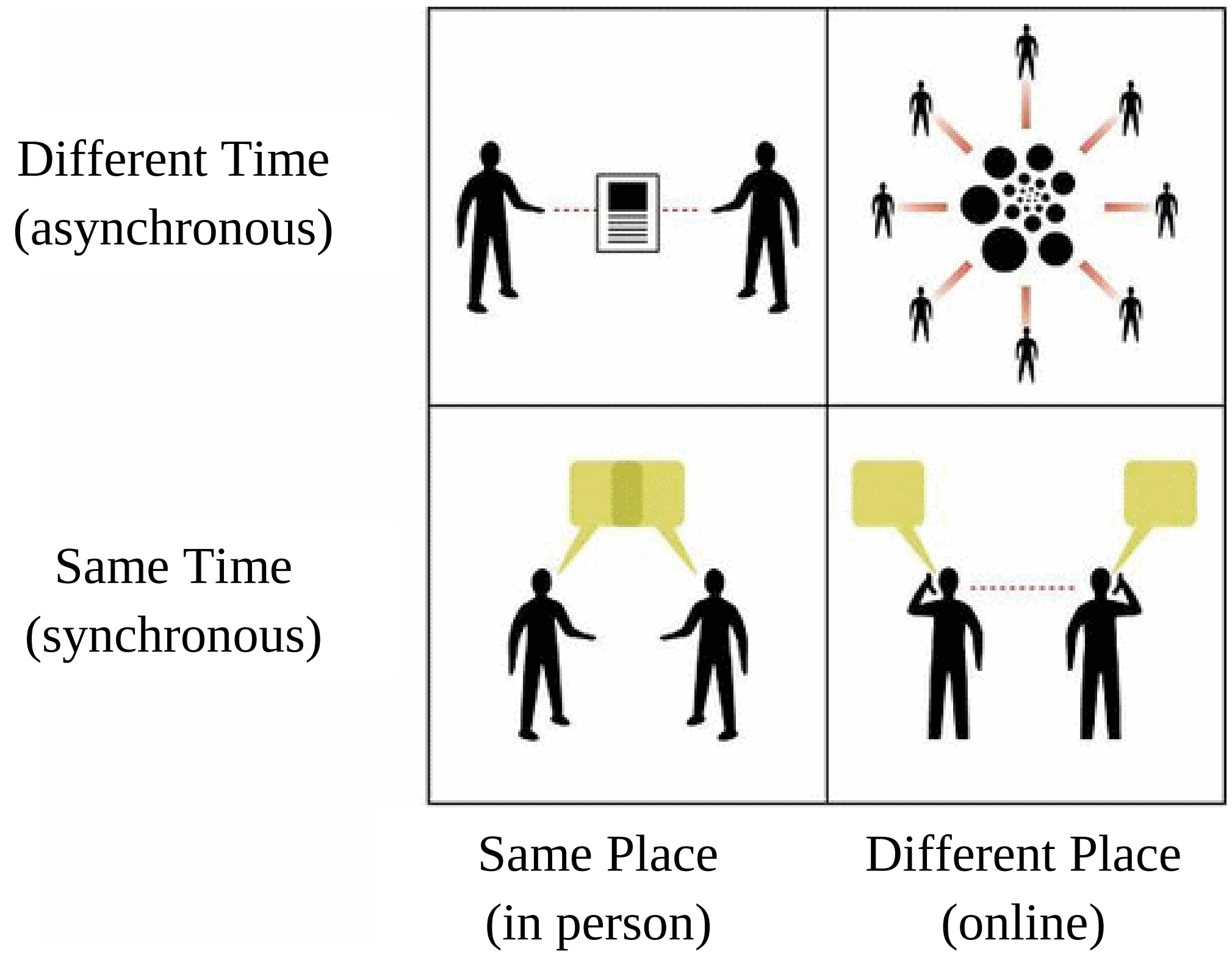

Figure 21. Communication ways [Swe13]

Within the aspect of *Communication* five main practices, which aim to build the successful communicational strategy when setting up global collaborations, were extracted from the literature – to define a Communication protocol and Team awareness channels, build Social relationships between collaboration sites, provide rich Communication tools and build up a Common knowledge base [SMS14].

### 4.6.1 Communication protocol

**Objectives:** The practice of defining a Communication protocol aims at identifying who is supposed to communicate with whom within the company – such as, to specify communication channels, interface points among teams and team members, sufficient frequency of communication, information exchange paths, and official corporate language [PaL03, Paa03, DeR09, SMS14].

**Potential results and experiences:** Defined and documented description of a communication protocol aims at creating awareness of team members from whom they will get work inputs and to whom they need to distribute work output results [SMS14]. Therefore it might ease coordination and reduce communication efforts, and make information flow more transparent and traceable [SMS14]. It aims to avoid delays in information exchange and minimize misunderstanding on roles and responsibilities between team members at different locations [PaL03, DeR09]. A communication protocol helps to create corporate organizational culture, helps to increase trust and make everyday software development work easier by having described communication rules [LeM07, NDC13, DeR09, Pyy03, SPF12, Dam07, KoO05, EDN01, MoS08, OKW08, RCM12, Paa03].

The study of Leszak and Meier on embedded product development in telecommunications domain in Alcatel-Lucent between Germany and China [LeM07], the report of Cusick and Prasad on the projects at Wolters Kluwer performed in the US and India [CuP06], the case study by Paasivaara and Lassenius based on interviews in 8 global software projects distributed across Europe, North America and Asia [PaL03], the study by Oshri et al. at LeCroy (Switzerland and USA), SAP (India and Germany), and Baan (India and The Netherlands) [OKW07], the research of Deshpande and Richardson based on interviews with 6 Irish software companies [DeR09] and the case study of Moe and Šmite in a Latvian software development company [MoS08] report that defining a communication protocol is an easy to conduct practice. Such practice brings no additional costs as documentation can be kept, updated and always available at project website or a wiki and saves time for finding a right person to contact and reduces misunderstanding. The absence of a communication protocol might lead to chaos during the information exchange, information overkill and wasting work time on forwarding data to the right person or finding the right contact person. Furthermore, it might result in delays for getting work inputs and distributing the outputs. Thus the absence of a communication protocol might result in losing some important project data.

**Actions:** Based on the organizational structure and peer-to-peer links define, document and distribute the communication protocol containing roles, responsibilities, tasks, organizational structure and communicational channels [LeM07, PaL03, DeR09]. Choose the official working language between collaboration sites (in most cases - English language) and provide language classes for team members if needed [EDN01]. Encourage communication between sites at different organizational levels whenever it is necessary without restrictions [MSL06].

### 4.6.2 Team awareness channels

**Objectives:** The practice of defining Team awareness channels aims at making collaboration team members become more familiar with remote colleagues and their skills, their expertise and availability, as well as their project activities and work status [SMS14].

**Potential results and experiences:** Based on the case study of Chang and Ehrlich in 3 global software development teams across the USA, UK and India [ChE07] four types of team awareness were identified – the awareness on team members' expertise and skills (How to find a contact person?), on problem solving practices (Whom to contact with a question?), on project current activities and project status, and on remote colleagues' availability. Creating channels for all types of team awareness is expected to build up teamness, trust and social ties between the collaboration sites. Clearly defined team awareness channels promise to achieve project visibility and transparency, to reduce delays for finding the right person to contact in case of questions, and to reduce time for problem solving [ChE07, PaL03, Paa03, PAD07, KoO05, SMS14].

**Actions:** Regarding the four types of team awareness in global collaborations different actions can be performed by organizations.

Based on the case study of Chang and Ehrlich in 3 global software development teams across the USA, UK and India [ChE07], the study of Bass et al. at Siemens [BHL07], the case study of Pyysiäinen on global projects in 9 Finnish companies operating in the domains of development of software products, bespoke systems and embedded systems [Pyy03], the study by Paasivaara and Lassenius based on interviews in 8 global software projects distributed across Europe, North America and Asia [PaL03], and the study of Kotlarsky and Oshri on two successful globally distributed system development projects at SAP (India and Germany) and LeCroy (Switzerland and USA) [KoO05] the team awareness on team members' expertise and skills can be created by using the organizational charts or project wiki/website/whiteboard that consist of contact information, team members' photos, roles and responsibilities. Such an approach "gives faces" to team members at different collaboration sites and helps to stimulate communication and teamness.

Based on the study of Paasivaara and Lassenius [PaL03] and the study of Kotlarsky and Oshri [KoO05] the team awareness on the problem solving practices could be created in three possible ways. For instance, an organization can establish the role of a Solution provider – a person with a lot of experience and expertise who could answer technical and other software system related questions. This action was successfully implied in two projects across Europe and Asia which were studied by Paasivaara and Lassenius [PaL03]. Another activity could be to establish Bulletin discussion boards or mailing lists that are dedicated to specific technology or a certain topic and could be used for posting and answering questions. Another promising approach could be to establish a "Problem e-mail box", a specific e-mail address where all the questions are sent to, then the moderator of this e-mail address is distributing questions to the person who has the required knowledge. It is expected to help in the beginning of the project when no personal contacts are yet established. The approach proved itself in a project across 6 sites in Europe studied by Paasivaara and Lassenius [PaL03].

Based on the case study of Prikladnicki et al. at 5 companies located in Brazil, Canada and the United States operating as a computer company, an IT service provider, a provider of enterprise energy management solutions and a retailing company [PAD07] and the study of Paasivaara and Lassenius [PaL03] it is recommended to create team awareness on project current activities and project status by using a project website that provides all the project related information including status and current activities, different groupware tools, weekly conference meetings, and an up-to-date always available project documentation that is located at project wiki/website.

Based on the study of Chang and Ehrlich in 3 global software development teams across the USA, UK and India [ChE07] and the case study of Niinimäki and Lassenius on global projects between Finland, Eastern Europe and India [NiL08] using of such tools as instant messaging that shows the availability status and shared calendars are expected to create the team awareness on remote colleagues' availability.

#### 4.6.3 Social relationships

**Objectives:** The practice of building Social relationships between collaboration sites represents the result of all communication activities and efforts [SMS14]. The practice aims at defining all sets of activities that need to be performed in order to build up and maintain the social ties between the collaboration sites at individual, team and organizational levels [OKW08].

**Potential results and experiences:** Social relationships between collaboration sites represent the base for building trust, common understanding, teamness and long-time successful collaboration in the end. Many studies report the importance of building and maintaining social interrelations between sites in global collaborations in order to achieve a successful way of working together. For example, the study of Oshri et al. on two successful globally distributed system development projects at SAP (India and Germany) and LeCroy (Switzerland and USA) [OKW08], the survey study of Hyysalo et al. [HPT06], the study of Leszak and Meier on embedded product development in telecommunications domain in Alcatel-Lucent between Germany and China [LeM07], the study of Hossain et al. [HBV09], the study of Pär J Ågerfalk et al. [AFH05], the study of Mettovaara et al. in Nokia and Philips [MSL06], the study of Paasivaara and Lassenius [PaL03], the case study of Casey and Richardson at 2 companies on a collaboration across the US, Ireland and the Far East in the financial sector [CaR05], and the case study of Ebert and De Neve at Alcatal operating in the telecommunications domain [EDN01]. Face-to-face meetings are reported as a highly efficient way for building social interrelations between collaboration sites [SMS14]. Even though face-to-face visits might cause additional investments and time, they need to be present, especially in the first phases of global collaborations [SMS14].

**Actions:** Based on the study of Oshri et al. on two successful globally distributed system development projects at SAP (India and Germany) and LeCroy (Switzerland and USA) [OKW08] it was identified that for building strong social ties between collaborations sites not only face-to-face meetings need to be performed, but also the activities before and after actual face-to-face meetings are rather important. Therefore three stages of building social relations can be distinguished: An introduction that represents the preparation activities before face-to-face meetings, a build up that represents the actual face-to-face sessions and building social relationships, and a renewal that represents the activities after face-to-face meetings for maintaining social ties (Figure 22). Figure 23 summarizes what activities could to be performed by organizations at each stage of building social relationships between collaboration sites regarding individual, team and organizational levels.

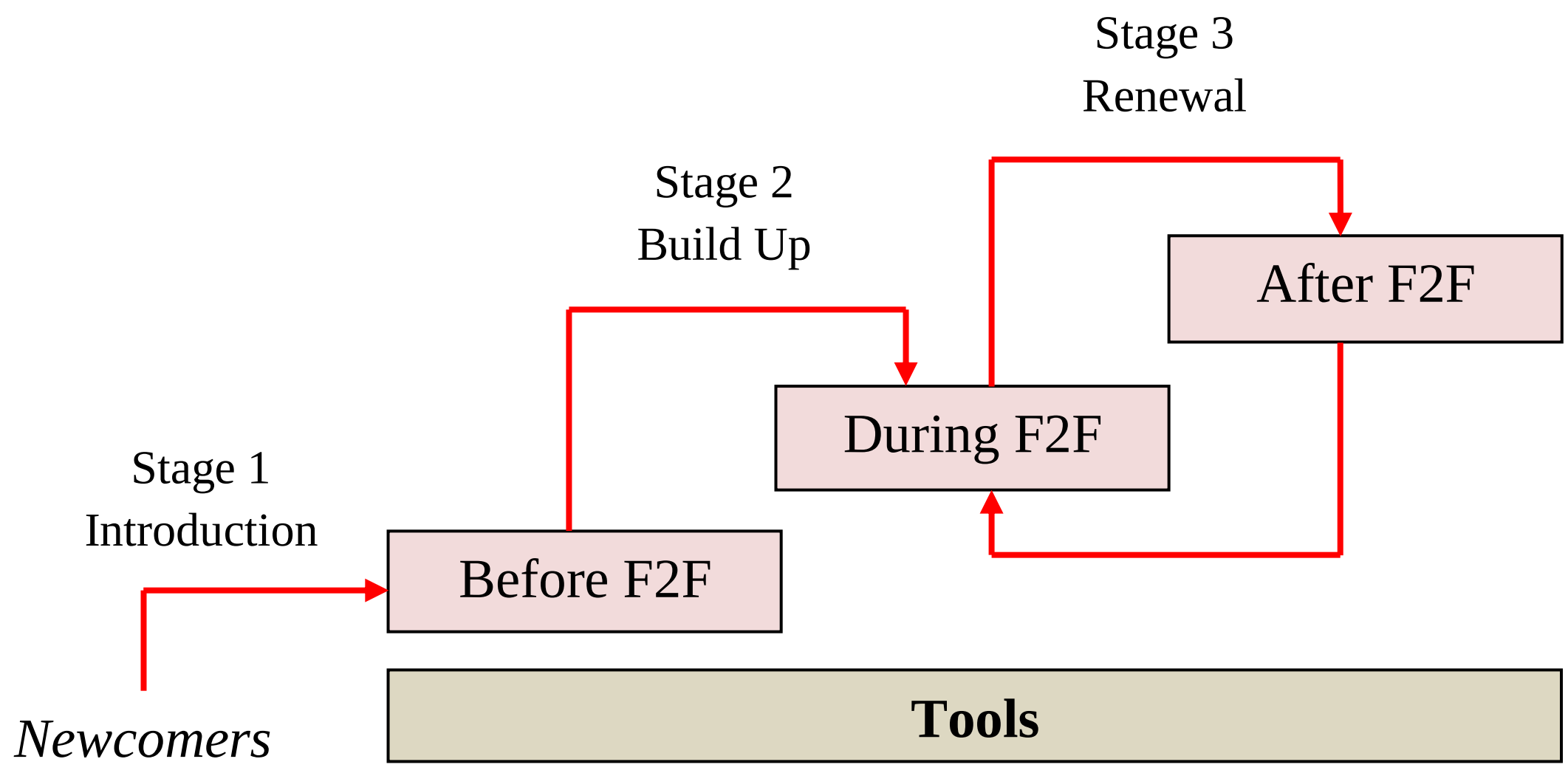

Figure 22. The stages of building up social ties [OKW08]

| | Introduction | Build-Up | Renewal |
|---|---|---|---|
| Individual | • Increase awareness of communication ways<br>• Offer language courses<br>• Offer short visits of individuals to remote locations | • Create space for one-to-one interactions<br>• Provide sense of importance to each member<br>• Adjust communication ways | • Ensure real-time communication channels<br>• Offer visits to remote locations<br>• Offer temporary co-location |
| Team | • Introduction of new team members<br>• Increase awareness of team composition<br>• Increase awareness of communication protocol<br>• Appoint contact person at each remote site<br>• Offer virtual F2F meetings | • Conduct kick-off meeting<br>• Discuss differences between national and organizational cultures<br>• Offer team-building exercises<br>• Organize social events<br>• Ensure awareness of organizational structure | • Facilitate reflection and feedback sessions<br>• Facilitate progress meetings<br>• Conduct virtual F2F meetings<br>• Offer F2F meetings |
| Organizational | • Distribute newsletters<br>• Create and offer shared cyberspaces | • Support sharing of information from f2F meetings (e.g. photos) | • Encourage direct communication channels |
| Tools | Phone, email, groupware tools, knowledge repositories, shared databases, teleconference, videoconference, online chat, intranet | | |

Figure 23. Activities supporting building social ties between collaboration sites [OKW08]

### 4.6.4 Communication tools

**Objectives:** Rich communication tools aim at supporting all the above-mentioned communication practices [SMS14]. Collaboration sites are often located at remote places, and the tools often provide the only way for software development teams to get connected [SMS14]. Thus a variety of different communication tools should be provided by organizations [TPB07, NPL10, OKW08, SMS14].

**Potential results and experiences:** The variety of communication tools helps to remote collaboration teams to communicate with each other. Often communication tools represent the only way of how distributed teams involved in global collaborations are able to communicate [SMS14]. Moreover, the tools stimulate communication between sites and make it more frequent that might help to achieve better common understanding, to ease coordination efforts and the way of working together, to build up trust, teamness and partnership feeling. Thus a variety of different communication tools such as web meetings, phones, e-mails and mailing lists, chats, file transfer tools, groupware and shared services tools should be provided by organizations to the GSD teams (Figure 24) [TPB07, NPL10, OKW08]. The case study of Thissen et al. at three distributed software teams from RTI International [TPB07] provides the examples of communication tools that could be used in global collaborations along with their advantages and infrastructure needs (Figure 25).

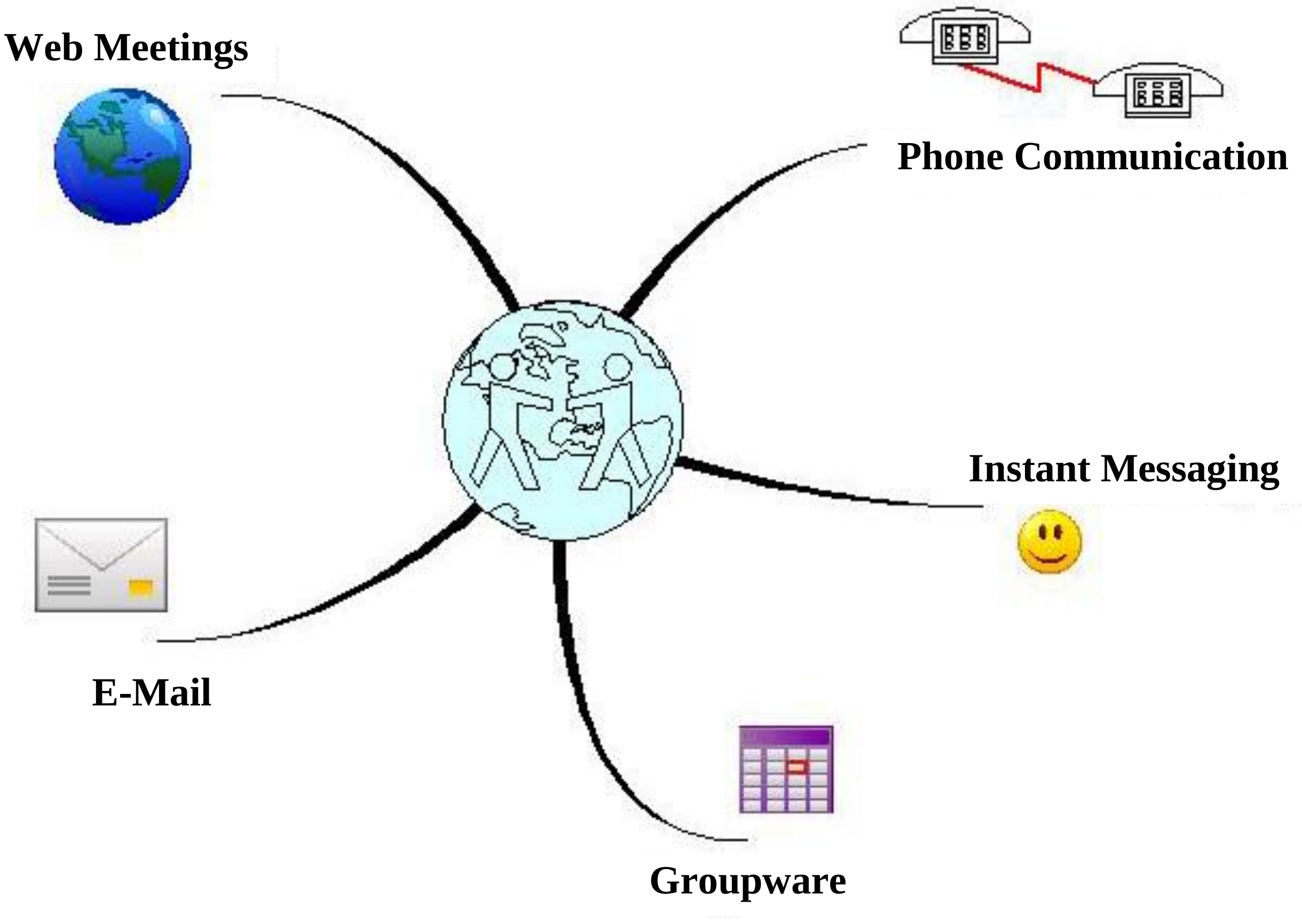

Figure 24. Communication tools to use by GSD teams [TPB07]

| Tool | Examples | Uses and Advantages | Bandwidth needs, Immediacy | Sensory Modes |
|---|---|---|---|---|
| Instant Messaging and Chat | • Yahoo messenger<br>• MSN messenger<br>• AOL Instant messenger<br>• IRC | • Instant interaction<br>• Less intrusive than a phone call<br>• View who is available<br>• Low cost<br>• Low setup effort | • Low bandwidth, can use dial-up<br>---<br>• Immediate<br>• Synchronous or asynchronous | • Visual<br>• Text and limited graphics |
| Groupware/ Shared services | • Lotus Notes<br>• Microsoft Exchange<br>• Novell GroupWise | • Calendars<br>• Contact Lists<br>• Arrange meetings<br>• Cost and setup effort vary | • High bandwidth is preferable<br>• Low bandwidth may be enough<br>---<br>• Asynchronous | • Visual |
| Remote Access and Control | • NetMeeting<br>• WebEx<br>• Remote Desktop<br>• pcAnywhere | • User controls a PC without being onsite<br>• Cost varies<br>• Setup varies | • High bandwidth<br>---<br>• Immediate<br>• Synchronous | • Visual<br>• Audio<br>• Tactile |
| Web Conferencing | • NetMeeting<br>• WebEx | • Live audio<br>• Dynamic video<br>• Whiteboard<br>• Application sharing<br>• Moderate cost and setup effort | • High bandwidth<br>---<br>• Immediate<br>• Synchronous | • Visual<br>• Unlimited graphics<br>• Optional audio |
| File Transfer | • File Transfer Protocol (FTP)<br>• Collaborative websites<br>• Intranets | • Share files on any type<br>• Cost varies<br>• Moderate setup effort | • Low bandwidth for small files<br>• High bandwidth for large files<br>---<br>• Asynchronous | • Varies with file content |
| Email | Numerous vendors and free applications | • Send messages or files<br>• Cost and setup effort vary | • Low bandwidth, can use dial-up<br>---<br>• Asynchronous | • Visual<br>• Audio in attached files |
| Telephone | • Plain Old Telephone Service (POTS)<br>• Voice Over Internet Protocol (VOIP) | • Direct calls<br>• Conference calls<br>• Cost varies<br>• Low setup effort | • VOIP requires high bandwidth<br>---<br>• Immediate<br>• Synchronous<br>• Asynchronous for voice mail | • Audio |

Figure 25. Examples of communication tools to be used in global collaborations [TPB07]

**Actions:** Provide a rich variety of communication tools to the GSD teams. Such as the tool examples that are provided above. Encourage communication between sites at different organizational levels whenever it is necessary without restrictions [MSL06].

### 4.6.5 Common knowledge base

**Objectives:** The practice of building up a Common knowledge base between collaboration sites aims at accumulating the collaboration experience in a specific organizational context. It aims at creating the "organizational memory", shared collective knowledge on the domain, technology, process, and business needs [SMS14, KoO05, Smi05, Rot06, SmW12, TPB07, SPF12, DeR09].

**Potential results and experiences:** Created common knowledge base between collaboration sites aims at easing the way of working together, increasing team awareness and mutual understanding. Based on the study of Kotlarsky and Oshri on two successful globally distributed system development projects at SAP (India and Germany) and LeCroy (Switzerland and USA) [KoO05] and the interview study of Rottman in outsourcing companies distributed in the US, Canada and India [Rot06] the "organizational memory" helps to create a corporate organizational culture shared by all team members and make global collaboration more partnership-oriented. Based on the case study of Šmite at a Latvian software company [Smi05] collective organizational experience and knowledge aim at helping collaboration sites to learn and improve their way of working together in the future projects.

**Actions:** Based on the interview study of Rottman in outsourcing companies distributed in the US, Canada and India [Rot06], the study of Šmite and Wohlin at Ericsson on the collaboration between a site in Sweden and a site in India [SmW12], the study of Höfner and Mani at an offshore software development center in India for a German organization [HoM07], the case study of Kobitzsch et al. at Tenovis GmbH &Co. KG operating in communication solutions domain on a collaboration between Germany and India [KRF01] and the case study of Moe and Šmite in a Latvian software development company [MoS08] co-located workshops, trainings and staff exchanges promise to create a common knowledge base between sites on the domain, development process, technology, understanding of the equipment, and business needs. Videotaping of trainings on domain knowledge and streaming them across collaboration sites are expected to support the common knowledge sharing in addition to co-located trainings. Project related websites/wikis are expected to store and accumulate the data on all the project activities [PAD07, PaL03].

## 4.7 Social aspects

The *Social aspects* address the processes through which team members gain knowledge on behavioral and communication norms, attitudes, cultural and social patterns of each other in order to work together in cooperation [OKW07, SMS14]. Global distribution of software development implies that individuals are usually not only geographically dispersed but also culturally [SMS14]. Thus the process of socialization and cultural integration is important when setting up global collaborations.

Socialization and cultural awareness between collaboration sites is expected to bring national and organizational cultures at remote locations together and create a mutual vision on the collaboration goals, needs and the whole process [OKW07, SMS14]. *"When there is a win-win situation the motivation is usually high and the chances of success get better"* [MSL06]. Socio-cultural awareness aims at creating the understanding of remote partners' way of working and behavior that promises to make global collaboration function successfully and beneficially for all involved sites [SMS14].

Socio-cultural distance might bring many challenges and negative effects into the collaboration process such as difficulties and inability of sites to communicate, unawareness of remote colleagues' qualification, unwillingness to exchange information, conflicts of tasks interpretation and unsuccessful end results [SMS14]. Those negative effects might have an impact on the end product quality and cause customer dissatisfaction [SMS14]. Therefore, organizations need to consider social aspects and stimulate socio-cultural integrity between collaboration sites [VBK14, SMS14].

Within *Social aspects* two main practices were extracted from the literature – to build Trust and to create Cultural understanding between collaboration sites [SMS14].

### 4.7.1 Trust

**Objectives:** The practice of building Trust aims at establishing effective, productive, reliable, and longitude collaborative social relationships between global software development teams [PNL12, Pyy03, SMS14]. Trust can be defined as the willingness of individuals to cooperate with others based on the belief that partners are reliable, competent and will do actions which are beneficial for the cooperation rather than for individual purposes [HoM07, SMS14].

**Potential results and experiences:** *"Trust is a pre-requisite for globally distributed software development"* [HoM07]. Many studies report the importance of building trust for the success of global collaborations and claim that trust needs to be built and maintained through the whole partnership history – from the first collaboration stages till its end [SMS14]. Such as the case studies of Ali Babar et al. with Vietnamese software developers working for Far Eastern, European and American clients [ABV07], the case study of Newell et al. on a collaboration across the US, Ireland and India in the financial sector [NDC07], the case studies of Piri et al. on global projects at 2 Scandinavian software companies for the projects in the financial sector, information and communication systems, enterprise resource management [PNL12], the case study of Moe and Šmite in a Latvian software development company [MoS08], the study of Oshri et al. at LeCroy (Switzerland and USA), SAP (India and Germany), and Baan (India and The Netherlands) [OKW07], and the case study of Casey and Richardson at 2 companies on a collaboration across the US, Ireland and the Far East in the financial sector [CaR05]. The studies conclude that trust promises to create the ability of remote collaboration sites to work together, and to build up the feeling of teamness and partnership [SMS14]. Trust stimulates the willingness of sites to communicate and work towards the completion of shared project goals – creation of **not "we and you"** relations **but "us"** [SMS14, PNL12, MoS08]. Moreover, trust is expected to increase the mutual flexibility and motivation of sites to collaborate, create higher tolerance on problems, stimulate productivity and achieve higher job satisfaction among employees [SMS14].

**Actions:** Building trust is a slow process, and a lot of efforts are needed to be done by organizations in order to build trust between collaboration sites. Three dimensions of building trust that need to be addressed by organizations were identified – Motivation, Communication and Socialization.

Based on the study of Kotlarsky and Oshri on two successful globally distributed system development projects at SAP (India and Germany) and LeCroy (Switzerland and USA) [KoO05], the interview study of Rottman in outsourcing companies distributed in the US, Canada and India [Rot06], the study of Mettovaara et al. in Nokia and Philips [MSL06] and the case studies of Piri et al. on global projects at 2 Scandinavian software companies for the projects in the financial sector, information and communication systems, enterprise resource management [PNL12] it is necessary to motivate teams at all collaboration sites. Thus motivate onsite teams through communicating the offshore strategy, targets, needs and potential outcomes – global collaboration is not a "job replacement", it is "do more with less resources". Motivate offshore teams in order to avoid staff turnover, treat offshore teams as an extension of onsite development teams. Organize face-to-face visits and stimulate frequent remote communication between collaboration sites via a rich variety of tools. Provide staff exchanges, socio-cultural trainings, social activities and relationships building in order to build social ties between collaboration sites and create trust [OKW07, MoS08].

### 4.7.2 Cultural understanding

**Objectives:** Cultural understanding represents shared norms and beliefs which are historically situated and followed by people belonging to a concrete society [SWG12, SMS14]. The practice of creating Cultural understanding aims at creating a mutual and shared understanding between collaboration sites on socio-cultural diversity of each other, especially with respect to the sense of time, social hierarchy, power distance, and preferable communication styles [SMS14].

**Potential results and experiences:** In the context of global collaborations socio-cultural diversity among sites might be interpreted as a facilitator for promoting creativity, innovativeness, and knowledge sharing [SMS14]. However, many studies report that at the same time cultural diversity might become a barrier for communication and effective coordination and therefore it needs to be addressed by companies when setting up global collaborations [SMS14]. Such studies as the case studies of Ali Babar et al. with Vietnamese software developers working for Far Eastern, European and American clients [ABV07], the research of Abraham on application of Hofstede matrix in software engineering [Abr09], the study of Huang et al. at a multinational information technology company [HuT08], the research of Deshpande and Richardson based on interviews with 6 Irish software companies [DeR09], the case study of Prikladnicki et al. at five companies located in Brazil, Canada and the United States operating as a computer company, IT service providers, a provider of enterprise energy management solutions, and a retailing company [PAD07], and the study of Mettovaara et al. in Nokia and Philips [MSL06] report the importance of culture-specific understanding and trainings that are aimed to reduce socio-cultural barriers between collaboration sites, establish one shared organizational culture, create mutual awareness, help to avoid conflicts and misinterpretations and ease the way of working together.

**Actions:** Build socio-cultural awareness between collaboration sites through stimulating the communication, establishing personal social ties, staff exchanges and trust building. Provide socio-cultural trainings on national and organizational cultures at onsite and offshore locations [Abr09, HuT08, KSW04, WDH08, LeM07, HoM07, LLA07, MHK05, JaM10, Hof14, OKW07, HCA06].

## 4.8 Infrastructure

The aspect *Infrastructure* refers to all tools, platforms and other technical means that support technical, organizational, and managerial activities in the context of distributed software development, maintenance, and operation [SMS14]. For instance, infrastructure includes tool support for coordination and communication between remote collaboration sites, IDEs, testing and quality assurance tools [SMS14].

Infrastructure is equally important for co-located and global software development environments. The global context implies additional and new requirements to the technological infrastructure that should be considered by organizations already in the first stages of collaborations [SMS14]. Thereby it is necessary for organizations to identify infrastructure-related requirements, to analyze the existing infrastructure, and to invest into the infrastructure in order to reach the stated requirements [VBK14, SMS14]. In addition, there is a need to analyze how the existing infrastructure at different sites can be modified so that it fits to a new distributed software development environment [SMS14].

With the aspect of *Infrastructure* two main practices were extracted from the literature – to provide Compatibility of infrastructure to all collaboration sites and to provide a rich variety of Tools for all the stages of software development [SMS14].

### 4.8.1 Compatibility

**Objectives:** The practice Compatibility aims at ensuring equal compatibility of technological infrastructure at all collaboration sites. For instance, collaboration sites should have equal internet connections, bandwidths, and communication facilities [VBK14, SMS14].

**Potential results and experiences:** A compatible infrastructure at all collaboration locations is highly important for conducting the distributed development process effectively and efficiently [VBK14, TPB07]. Compatibility is important, for instance, for configuration management environments, for development tools, and for coordination support [SMS14]. The compatible infrastructure at all collaboration sites makes the software development process faster and easier to perform, eases coordination efforts and helps to achieve high productivity in task completions at all locations.

**Actions:** Provide equal compatible technological infrastructure at all locations involved in the global collaboration such as internet connection, bandwidth, development environments and rich communication infrastructure (i.e. tools, video/phone conference rooms).

### 4.8.2 Tools

**Objectives:** The practice Tools aims at providing a rich variety of tools to all collaboration sites for performing successfully all tasks for software product development. Tools include, for instance, coordination and communication tools, IDEs, different groupware tools, testing and quality assurance tools [TPB07, SPF12, SMS14].

**Potential results and experiences:** A rich variety of tools in global software development projects aims at mitigating the impact of distance in GSD, allows development teams at different locations successfully perform software development tasks in an effective way and supports the completion of collaboration goals [SMS14]. For instance, coordination tools and communication tools promise to help mitigate communication risks that are due to temporal, geographical, and cultural distances [SMS14]. A rich set of groupware tools is expected to help reduce the impact of distance in global software development, to increase the frequency and ease the communication between sites, to lessen coordination efforts, and to provide equal accessibility to all project-related artefacts [TPB07, SPF12, VBK14, SMS14].

**Actions:** Provide a rich variety of tools to all locations involved in the global collaboration such as coordination and communication tools, IDEs, different groupware tools, testing and quality assurance tools.

## 4.9 Organizational change process

The aspect *Organizational change process* refers to all activities and process changes that organizations need to perform when doing a transition from co-located software development into a global context. When organizations start setting up global software development collaborations, there is clear evidence that a sufficient amount of time is needed in order to gain desired efficiency [SmW12, SMS14]. At first, challenges such as communication, coordination, trust building, team awareness and integration of working procedures lead to significant reductions of the overall work process efficiency [SMS14]. Reasons for the decrease in the work efficiency during the first collaboration stages are usually the time that is needed for building a compatible infrastructure, establishing the necessary communication paths between collaboration sites, providing domain, technology and cultural training, as well as building social relationship and teamness [SMS14]. After the decrease in productivity and work efficiency during the first stages of global collaborations, there is typically a period of time “for the recovery” when partners learn to know each other and better understand the ways of working together in cooperation [SMS14]. In this period, the software development efficiency and productivity are usually recovering steadily [SMS14]. After this phase, global collaborations might get scaling effects and reach the efficiency that is higher than the efficiency of a co-located development [SmW12, MoW01, BBA13, SMS14]. Figure 26 summarizes the phases and challenges of transition from co-located to global software development.

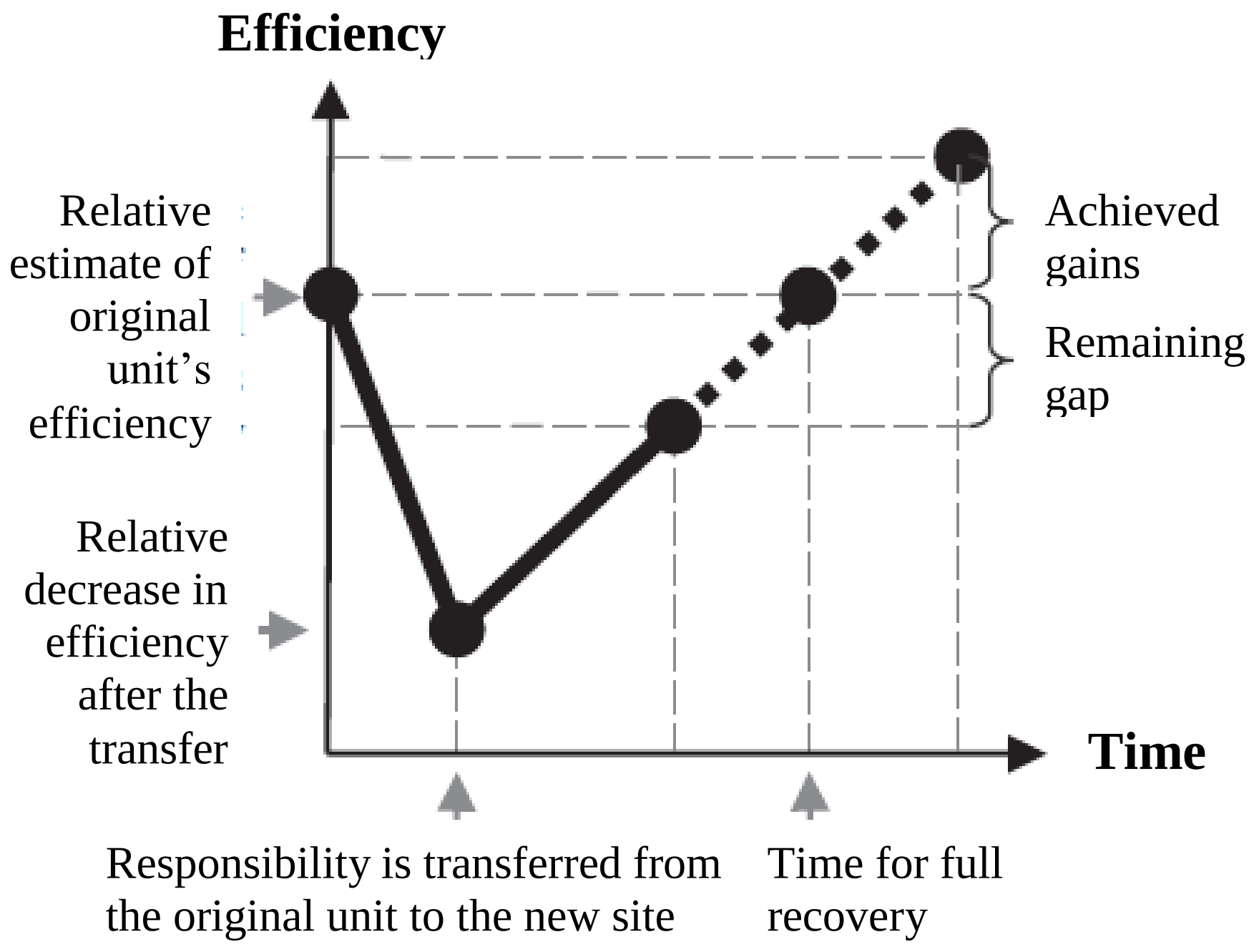

Figure 26. The challenges of transfer from co-located to global software development [SmW12]

Gaining the scaling effects and high level of efficiency in global collaborations requires the change in organizational working processes and the establishment of systematic process improvement procedures [SMS14].

Within the aspect of the *Organizational change process* two main practices, which need to be considered by organizations, were extracted from the literature – to perform process Improvement cycles and to establish a collaboration process Standardization [SMS14].

### 4.9.1 Improvement cycles

**Objectives:** The practice of performing process Improvement cycles refers to all activities that aim at accumulating organizational collaboration working history with respect to the transfer from co-located software development into a GSD working style, changing software development processes and performing improvement actions [BBA13, Smi05, SmW12].

**Potential results and experiences:** The accumulated collaboration working experience in specific organizational context aims at creating shared "organizational memory" including domain and business needs, technical aspects, work process aspects and decision history [Smi05]. Such organizational knowledge gives a lot of insights regarding the setting up of global collaborations and should be iteratively examined by organizations in order to achieve process improvements, increase productivity, end product quality and benefit even more when setting up global collaborations [SMS14]. Improvement cycles aim at learning and improving the whole process of working in cooperation.

**Actions:** Discuss and analyze new ideas for process changes and improvement actions jointly by the collaboration sites on a regular basis during the whole period of the collaboration.

The improvement of a globally distributed collaboration can follow different process improvement approaches. Such as, the CIP or continuous improvement process that is designed to review and evaluate work policies and procedures continuously in order to find strategies for enhancing the overall productivity and efficiency [Nic14]. Another approach could be the Kaizen lean management improvement methodology that is aimed at eliminating waste, improving productivity, and achieving sustained continual improvement within organizational processes [EPA11]. The Kaizen method includes problems definition and analysis, preparation of improvement plan, its implementation, measuring and the gained results to the targets. TQM or Total Quality Management approach is a process measurement based quality-oriented approach that is used by organizations to improve their internal processes, overall performance and increase customer satisfaction [Rey13].

However, there is a lack of experience in improvement approaches that are focused on global collaborations [SMS14]. Therefore, it is recommended to deploy a problem-oriented, continuous improvement approach that combines a few process improvement methodologies [SMS14].

### 4.9.2 Standardization

**Objectives:** The practice Standardization aims at achieving a high level of standardization of the overall global software development process. This includes, for instance, creating standardized corporate culture, communication and coordination practices, software development approaches.

**Potential results and experiences:** A high level of process standardization aims at easing the overall operation of global collaboration process, reducing coordination and management efforts, increasing work productivity and efficiency. Standardization is a result of continuous process improvement that aims in a high level at improving end product quality and customer satisfaction [SMS14].

**Actions:** Perform continuous process improvement actions and establish standardized norms and processes within the organization. Monitor the results continuously and perform changes if necessary.

# 5 Global Canvas

## 5.1 Introduction

When setting up global collaborations in the domain software development organizations need to consider different aspects, practices, strategies and phases of collaborations. A holistic approach which aggregates existing experiences from the academia and industry regarding establishing global collaborations is widely missing [SMS14]. The literature review that was performed in this thesis and the close collaboration with the case company resulted in extracting principal aspects and successful practices that need to be addressed by companies when doing global software development collaborations. There is a need to present defined aspects and practices in an attractive, relevant, credible, useful way that could be helpful for the practitioners. Results from the literature study were structured in the form of a worksheet, the so-called “Global canvas”, that presents nine main aspects that need to be addressed by organizations when doing global collaborations. Each aspect of the worksheet consists of successful practices that need to be performed by companies when setting up global collaborations. The practices in the worksheet are prioritized in the form of activity roadmaps that could be used by the industry as a guide for setting up global software development collaborations. The initial proposal of the worksheet “Global canvas” is described in this section.

## 5.2 Elements

The worksheet “Global canvas” that is proposed to be a practical guide for establishing global collaborations in the software development domain includes three main elements. These elements are called *Aspects*, *Practices*, and *Phases* [SMS14].

The element *Aspects* represents principal aspects that need to be addressed by companies when establishing global collaborations in software development domain. “Global canvas” consists of nine principal aspects that are presented in the Figure 27 [SMS14].

The element *Practices* represents for each defined aspect a set of successful practices that need to be performed by organizations when setting up global collaborations. Figure 28 presents the full set of extracted practices and how they are grouped with respect to aspects of global software development collaborations [SMS14].

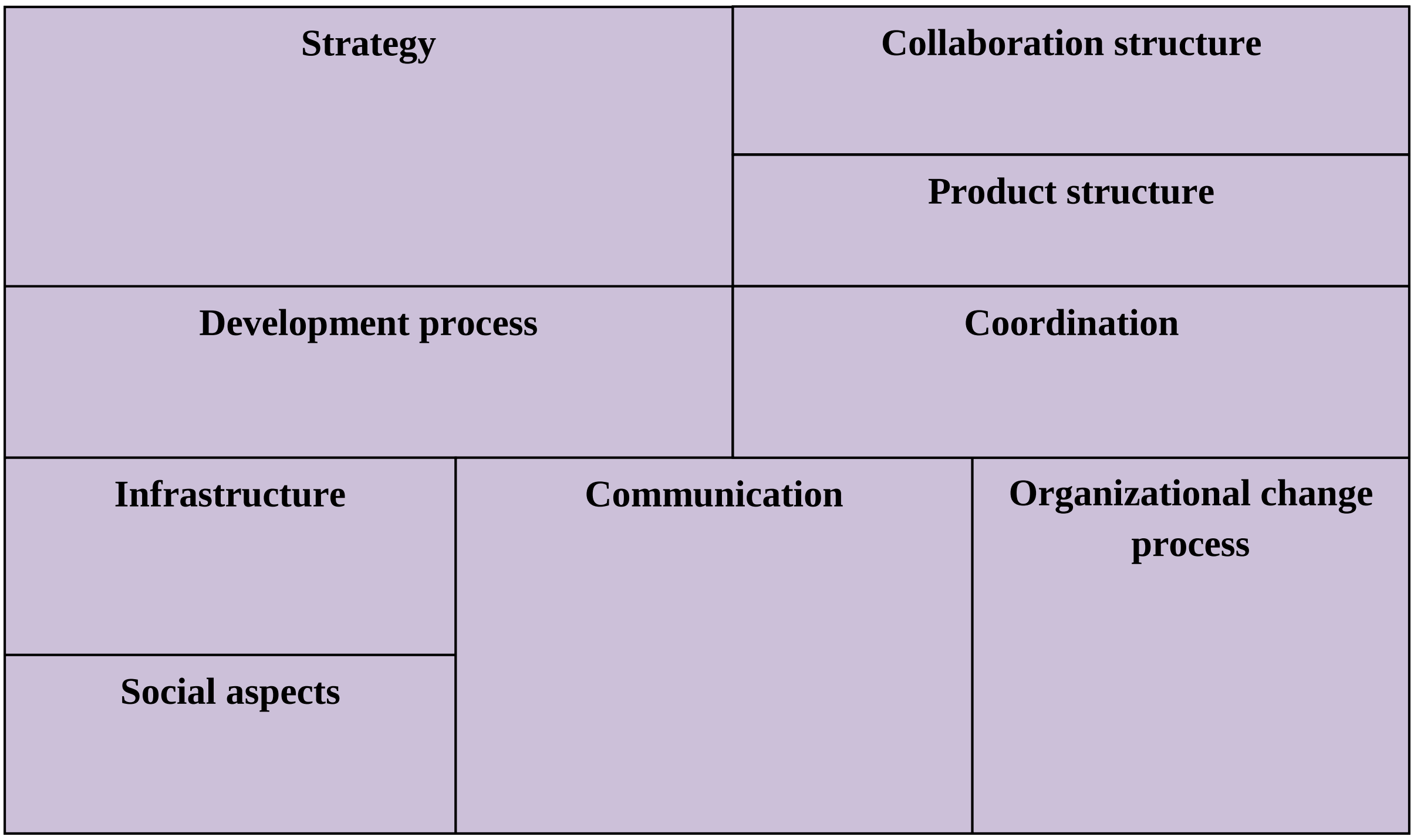

Figure 27. Aspects of global software development collaborations (based on [SMS14])

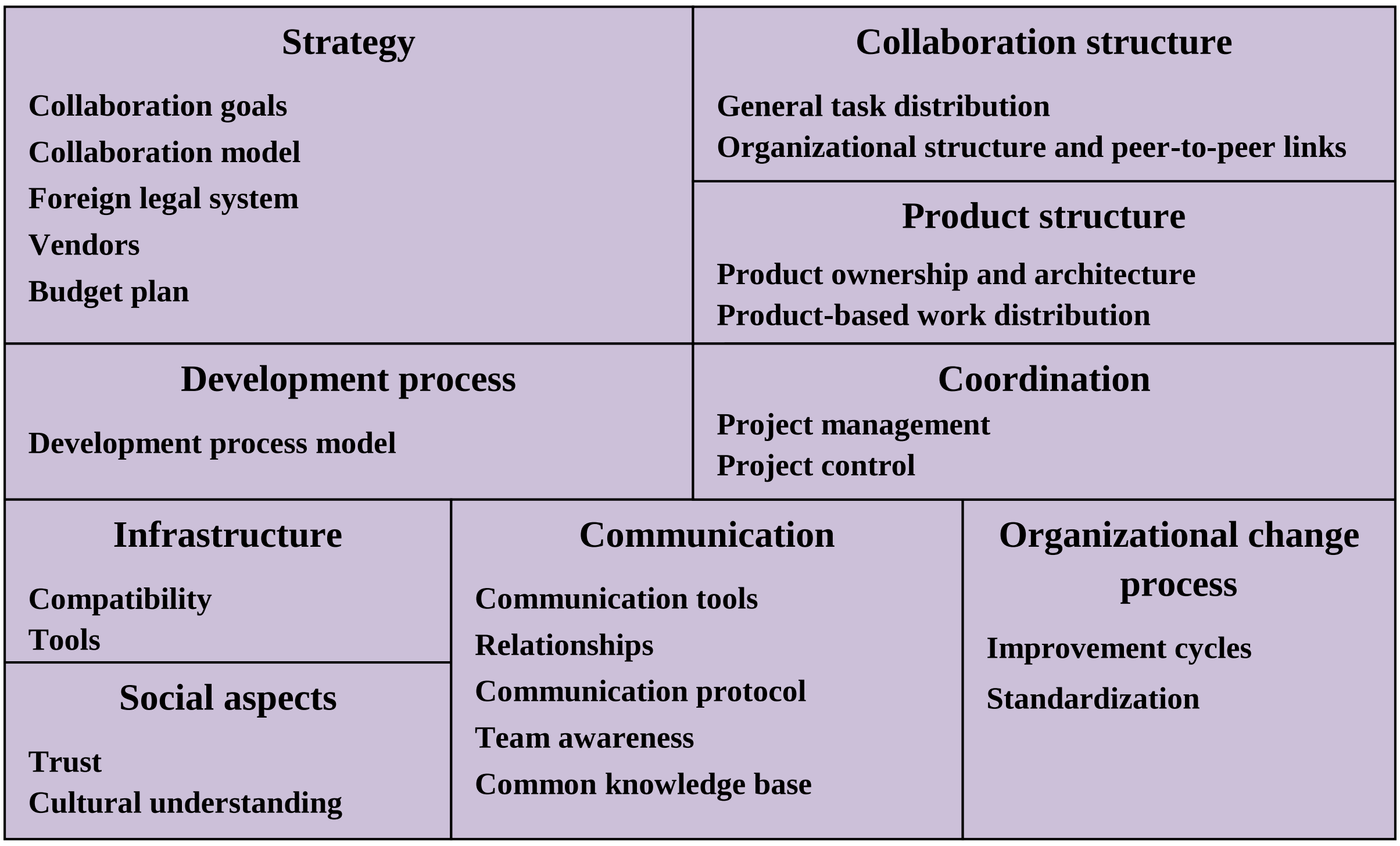

Figure 28. Practices for each aspect of global software development collaborations (based on [SMS14])

The element *Phases* represents the phases that could be distinguished when setting up global collaborations. Each collaboration phase can be characterized by a specific set of activities that need to be done by companies, in order to establish successful global software development collaborations. Four main phases of global collaborations were distinguished – Initiate, Plan and Prepare, Pilot, and Operate and Improve [SMS14].

Based on defined collaboration phases, principal aspects and practices that need to be addressed when setting up global collaborations were structured in the form of activity roadmaps. The initial sequence of activities was provided by the case company and refined at a joint workshop of Daimler and the University of Helsinki [SMS14]. The final order of activities was created mainly based on the experiences reported by project leaders from the case company and results from the literature study. Some relations between activities also have an underlying inner logic. The sequence of activities is not a strict order and can be customized for specific organizational needs. The activity roadmaps are visualized through the worksheet "Global canvas" that can be adjusted for specific organizational context and used as a guide for companies in systematically setting up global collaborations for software-based products and services [SMS14]. The proposed worksheet is explained in the following chapter.

## 5.3 "Global canvas" visualization

This section describes the proposed worksheet "Global canvas" (Figure 29) that presents the activity roadmaps for organizations intending to set-up global software development collaborations [SMS14].

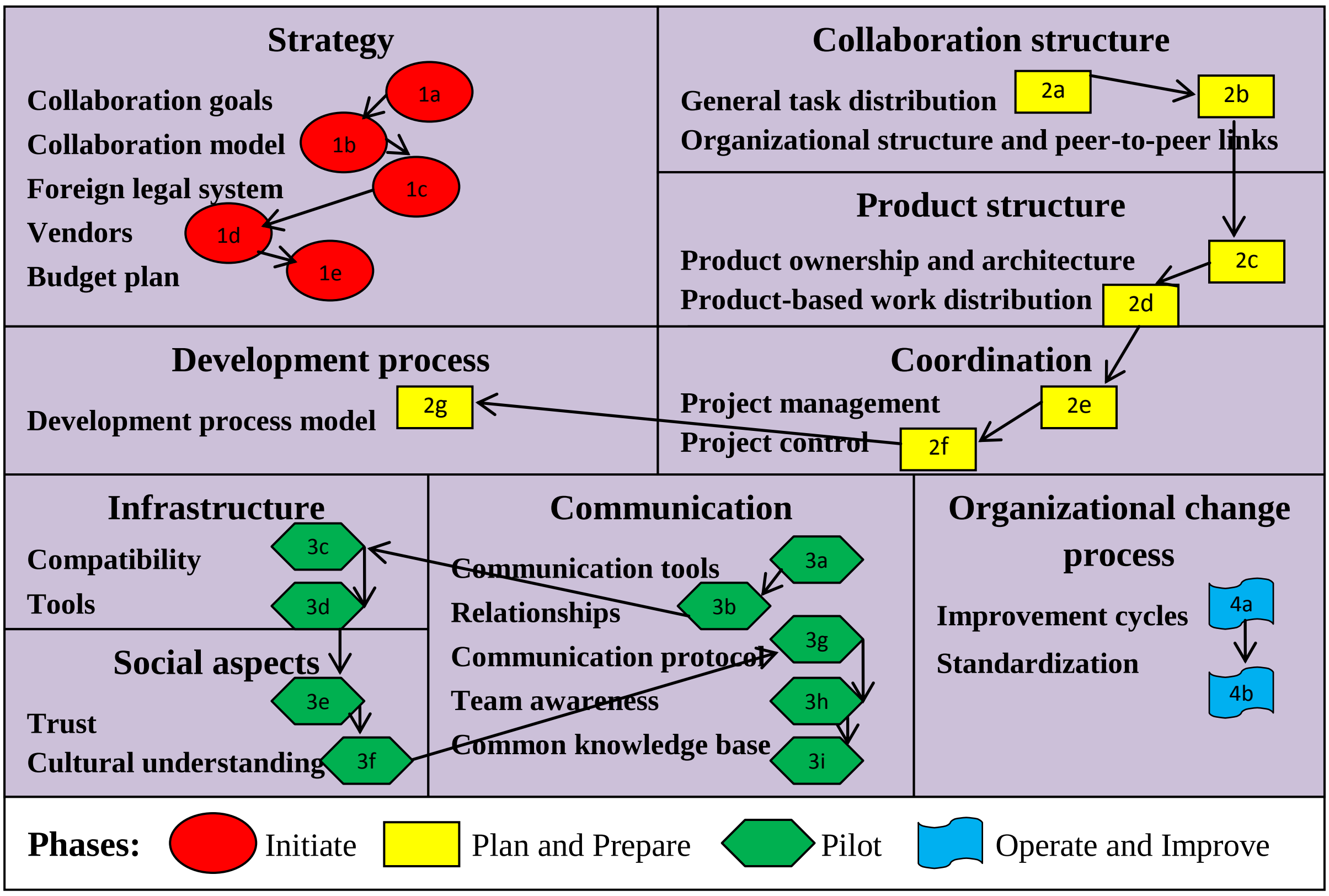

Figure 29. "Global canvas" guidance worksheet [SMS14]

Phase 1. **Initiate**.

This is the initial stage that organizations face when deciding to transfer software development activities from co-located into a global context and set-up global collaborations [SMS14]. In this phase organizations need to investigate the potential results that the transition into a GSD brings to the organizational context, including possible benefits and challenges. Companies need to analyze the existing models of global collaborations and choose what model suits their specific organizational environment better. Furthermore, in this phase organizations make first decisions regarding potential collaboration partners. Thus, the proposed sequence of activities to be addressed by the organization at the initiation phase might look like the following [SMS14]:

a) Identify needs and goals for doing a global software development collaboration. Analyze expected outcomes and benefits that global collaboration might bring to the specific organizational context.

b) Choose a global collaboration model that is suited for the specific organizational context and the business needs.

c) Investigate the foreign legal system(s) concerning IP and contract laws.

d)Choose appropriate partner(s)/vendor(s) with sufficient infrastructure, capabilities and expertise needed for the chosen collaboration model.

e) Define a budget plan for doing global software development projects. Include possible hidden costs such as communication tools, face-to-face visits or infrastructure needs.

Phase 2. **Plan and Prepare**.

This phase of global collaborations can be considered as a preparation stage. The phase aims at building all the conditions needed for global collaborations to start fully functioning [SMs14]. Organizations in this phase define task distribution and suitable product structure, divide roles and responsibilities between sites, choose coordination mechanisms and development process model [SMS14]. In this phase, the organization still keeps ongoing product development mainly onsite [SMS14]. However, at the same time, the organization pilots first practices of setting up global software development [SMS14]. The proposed sequence of activities to be addressed by the organization at the planning and preparation phase might look like the following [SMS14]:

a) Identify general task distribution based on a development process breakdown.

b) Define and document the organizational structure including specific roles, responsibilities and peer-to-peer links between collaboration sites.

c) Identify/adapt the model for product architecture and ownership boundaries between sites (based on the collaboration model defined earlier and the development process breakdown).

d) Define a product-based work distribution between sites based on available resources and capabilities.

e) Define the coordination mechanisms between collaboration sites. Choose an appropriate project management model that suits the chosen collaboration model.

f) Specify project control procedures for monitoring work progress and detecting problems as early as possible.

g) Choose a model according to which the software development process will be performed.

Phase 3. **Pilot**.

This phase of global collaborations aims at systematic testing of practices necessary to be addressed when setting up collaborations [SMS14]. In this phase it is possible to detect the problems of collaboration and react to them early on [SMS14]. If some things do not work at first they can be changed [SMS14]. The proposed sequence of activities to be addressed by the organization at the piloting phase might look like the following [SMS14]:

a) Provide a rich variety of communication tools in order to stimulate communication between sites and to avoid misunderstandings.

b) Stimulate building social relationships between collaboration sites (e.g., organize face-to-face visits, joint social activities, staff exchanges, stimulate communication).

c) Ensure that the remote partner(s) has sufficient infrastructure needed for software development projects. Provide compatibility of internet connections, bandwidths, communication facilities (for instance, video conference rooms) between sites.

d) Introduce a rich variety of groupware tools that are aimed to ease the collaboration process between sites.

e) Consider the impact of socio-cultural distance between partners. Start building trust between sites early on.

f) Ensure cultural awareness between sites regarding different cultural perceptions that might occur in collaboration between partners belonging to distinct societies.

g) Establish communication protocol. Identify who should communicate with whom and how often. Make team members understand that communication is an important part of everyday work.

h) Ensure team awareness channels. Team members need to be aware of remote colleagues' contact details, expertise, roles and responsibilities, work schedules. Ensure that the teams are aware of the project tasks and statuses.

i) Accumulate the collaboration experience based on the working history between sites, and create a collective shared knowledge base – the "organizational memory".

Phase 4. **Operate and Improve**.

In this phase the overall operation of a global software development process is ongoing [SMS14]. Based on the accumulated working history, the collaboration partners learn better how to work together, propose and handle process changes and improvements [SMS14]. This phase aims at achieving a continuous improvement of the global collaboration, in order to increase the efficiency and effectiveness. The proposed sequence of activities to be addressed by the organization at the operating phase might look like the following [SMS14]:

a) Analyze the working history between sites. Continuously discuss potential process changes and improvements. Monitor the results of improvement actions.

b) Improve the process of global collaboration continuously and thereby aim at achieving a high level of process standardization.

## 5.4 "Global canvas" use cases

The developed worksheet "Global canvas" provides a holistic "shopping list" of aspects and practices that need to be addressed by companies when doing global collaborations [SMS14]. The aspects and practices are prioritized in the form of activity roadmaps that makes the worksheet to be a feasible guide that can be used in the industry when establishing global collaborations [SMS14]. Furthermore, the canvas can be seen as a reminder of all major activities that are suggested to be performed in order to set-up successful, longitude global software development collaborations. In the mid-term perspective the canvas can be also used as an assessment scheme for already established collaborations [SMS14]. The future research plans include the analysis of the canvas' suitability for other purposes than being a guide and an assessment scheme.

# 6 Discussion and Limitations

Based on the performed literature review and advices from the industrial case company, this thesis aimed at investigating and aggregating existing experience on what aspects and practices need to be considered by practitioners when setting up global collaborations. The defined aspects and practices were grouped together and prioritized in the form of activity roadmaps. The initial version of a holistic worksheet “Global canvas” that shows the visualization of the activity roadmaps was presented. The canvas is proposed to be a guidance for companies intending to set-up long-term global collaborations in the software development domain [SMS14].

However, the thesis includes a set of threats to its validity that need to be discussed. Following Shadish et al. [SCC02], the different types of validity are threatened in this study:

- Conclusion validity refers whether the results of the study are statistically proven and significant. This study was performed as a qualitative study. The results, particularly the canvas and prioritization of aspects and practices, are based on the analysis of literature findings and industry inputs. Thus the conclusion validity is quite low.
- Internal validity refers whether the results are based on causal relationship. For most of the defined aspects and practices that need be addressed when doing global collaborations the literature discusses the potential outcomes and challenges. Therefore, the results are quite likely to reflect the reality and cause-effect relationship of the aspects and practices on the GSD collaborations. However, the selection of the principal aspects and practices and their inclusion into the canvas was made on assumptions that imposes a thread to internal validity of the developed guide.
- Construct validity refers whether the measured samples highly match research constructs and real-world situations. For this thesis study the literature review was performed as a main research method. Thereby, the practices extracted from the literature, especially from the case studies, represent the real-world experiences. Furthermore, the collaboration with the case company and continuous feedback from them based on their experience makes the results real-world oriented and gives an initial study validation. Thus, the construct validity of this thesis study might be considered being relatively high.
- External validity refers whether the study results can be generalized. The canvas and sequence of activities for doing global collaborations is not a strict order and mainly based on assumptions, literature findings, and industry inputs. The presented aspects derived from the case company experience might differ for other contexts. Thus, the generalization of results is quite limited and requires more experience and testing in the industrial context.

# 7 Conclusions and Future Work

The main goal of this thesis was to create a holistic approach for setting up global collaborations which aggregates aspects, practices, strategies, and existing experiences and guides companies in systematically establishing global collaborations for software-based products and services. Such a holistic approach should be easy to use, practical, credible and relevant for practitioners intending to start global software development collaborations. A literature review following the Snowballing approach was performed as a main research method. In addition, there was collaboration with an industrial partner that became a case company for this thesis study. The achieved results and new contribution of this thesis can be summarized under the following aspects.

The study investigated and extracted a “shopping list” of aspects and practices that are necessary to be addressed by organizations when setting up global collaborations. The defined aspects and practices were prioritized and structured in the form of activity roadmaps. As a visualization of such activity roadmaps, an initial proposal of a worksheet was developed, the so-called “Global canvas”, that presents scientific experiences and findings in an effective way and guides companies in systematically setting up global software development collaborations. The canvas aims at being not only a guide, but a reminder of all principal activities and things that companies need to think of when doing global collaborations. Furthermore, the canvas is proposed to be as an assessment scheme for already established collaborations.

After the completion of this thesis, a few questions still remained unanswered that need be investigated in the future work. First of all, the mature and complete validation of the developed worksheet needs to be studied in detail as it was out of scope in the thesis. The canvas and its general applicability need more systematic testing in the industrial context. The dependencies between different practices and strategies, as well as other potential use cases of the canvas, are also planned to be analyzed.